\documentclass[prb, superscriptaddress, twocolumn, preprintnumbers,showpacs]{revtex4}
\usepackage{ifpdf}
\usepackage{graphicx}
\usepackage{amsmath}
\usepackage{multirow}
\usepackage{bm}
\usepackage{color}
\usepackage{ulem}

\begin{document}
\normalem
\title{Broken-symmetry $\nu=0$ quantum Hall states in bilayer graphene: \\
Landau level mixing and dynamical screening}
\date{\today}


\author{E.V. Gorbar}
\email{gorbar@bitp.kiev.ua}
\affiliation{Department of Physics, Taras Shevchenko National Kiev University, 03022, Kiev, Ukraine}
\affiliation{Bogolyubov Institute for Theoretical Physics, 03680, Kiev, Ukraine}

\author{V.P. Gusynin}
\email{vgusynin@bitp.kiev.ua}
\affiliation{Bogolyubov Institute for Theoretical Physics, 03680, Kiev, Ukraine}

\author{V.A. Miransky}
\email{vmiransk@uwo.ca}
\affiliation{Department of Applied Mathematics, University of Western Ontario,
London, Ontario N6A 5B7, Canada}

\author{I.A. Shovkovy}
\email{igor.shovkovy@asu.edu}
\affiliation{Department of Applied Sciences and Mathematics, Arizona State
University, Mesa, Arizona 85212, USA}

\begin{abstract}
For bilayer graphene in a magnetic field at the neutral point, we derive and solve a full set of gap
equations including all Landau levels and taking into account the dynamically screened Coulomb
interaction. There are two types of the solutions for the filling factor $\nu=0$: (i) a spin-polarized
type solution, which is the ground state at small values of perpendicular electric field $E_{\perp}$,
and (ii) a layer-polarized solution, which is the ground state at large values of $E_{\perp}$.
The critical value of $E_{\perp}$ that determines the transition point is a linear function of the
magnetic field, i.e., $E_{\perp,{\rm cr}}=E_{\perp}^{\rm off}+a B$, where $E_{\perp}^{\rm off}$ is
the offset electric field and $a$ is the slope. The offset electric field and energy gaps
substantially increase with the inclusion of dynamical screening compared to the case of
static screening. The obtained values for the offset and the energy gaps are comparable
with experimental ones.
The interaction with dynamical screening can be strong enough for reordering the
levels in the quasiparticle spectrum (the $n=2$ Landau level sinks below the $n=0$ and $n=1$ ones).
\end{abstract}


\maketitle

\section{Introduction}
\label{Introduction}

Bilayer graphene is a unique material in condensed matter physics. It combines some characteristics of monolayer
graphene and more traditional two-dimensional electron systems. Its low energy electron spectrum is gapless and
is given by parabolic valence and conduction bands with massive chiral charge carriers touching at two $K$ and
$K^{\prime}$ valley points. As was first suggested in Ref.~\onlinecite{Falko} and experimentally shown in
Ref.~\onlinecite{Ohta}, if a potential difference between the layers is applied, a tunable energy gap is opened
at the valley points. In fact, as was indicated in Ref.~\onlinecite{Zhang},
even without an external electric field, the quadratic dispersion relation in bilayer graphene implies that the
electron-electron interaction should open a gap in the spectrum at the neutral point in clean bilayer samples
(for a similar development in the theory of topological insulators, see Ref.~\onlinecite{Sun}).

The rich spin-valley approximate $SU(4)$ symmetry of the low energy electron Hamiltonian with the Coulomb
interaction leads to many interesting possibilities for the gap generation. In the absence of magnetic field,
anomalous quantum Hall (QAH),\cite{Levitov-anomalous,Fzhang} quantum spin Hall (QSH), layer
antiferromagnet (LAF), and nematic \cite{Vafek,Lemonik} states were suggested as possible ground
states of bilayer graphene at the neutral point (for a general discussion, see Ref.~\onlinecite{MacDonald}).

As is well known, a magnetic field is a strong catalyst of symmetry breaking in graphene like
systems.\cite{catal,Khvesh,graphite}
In the presence of a magnetic field, the gap generation was experimentally
observed in Refs.~\onlinecite{Feldman,Zhao,Weitz,Martin,Kim,Freitag,Velasco,Elferen}. It was found that the
eightfold degeneracy in the zero-energy Landau level can be lifted completely, giving rise to
the quantum Hall states with filling factors $\nu=0,\pm 1,\pm 2, \pm 3$. In suspended bilayer
graphene used in Refs.~\onlinecite{Feldman,Weitz,Martin}, the values of the gaps are of the
order of a few milli electron volt for magnetic fields $B\sim 1\,\mbox{T}$.
The theory of the quantum Hall (QH) effect in bilayer graphene has been studied in
Refs.~\onlinecite{Barlas,Abergel,Shizuya,Nakamura,Nandkishore1,GGM1,GGM2,Nandkishore2,Toke,Kharitonov1,GGJM}.

It was revealed in Ref.~\onlinecite{Feldman} that the energy gaps scale linearly with magnetic field $B$
in bilayer graphene. This is in contrast to the case of monolayer graphene where a $\sqrt{B}$
scaling for the gaps takes place.\cite{Jiang2007} As was suggested in Refs.~\onlinecite{Nandkishore1,GGM1,GGM2},
a strong screening of the Coulomb interaction is responsible for this modification of the
scaling in bilayer. The physics underlying this effect is the following.\cite{GGM1,GGM2}
The polarization function in bilayer graphene is enhanced much more than in monolayer graphene 
and, as a result, its contribution to the effective interaction dominates over that of the bare Coulomb 
potential for magnetic fields $B \lesssim 30\,\mbox{T}$.\cite{GGM2} Such a strong screening radically 
changes the form of the interaction and leads to a linear scaling law.

Another interesting phenomenon in the $\nu = 0$ QH state in bilayer graphene is the phase transition
between the spin polarized (ferromagnetic) phase and the layer polarized one in the $B$-$E_{\perp}$ plane,
where $E_{\perp}$ is an electric field orthogonal to the bilayer planes. It was predicted in theoretical studies
in Refs.~\onlinecite{GGM1,GGM2,Nandkishore2,Toke} and observed in experiments in
Refs.~\onlinecite{Weitz,Kim,Velasco}.

In suspended bilayer graphene used in Refs.~\onlinecite{Feldman,Weitz,Martin}, the mobility is
in the range from $10,000$ to $15,000\,\mbox{cm}^2/\mbox{Vs}$ and the values of the gaps are of the
order of $1\,\mbox{meV}$ for magnetic fields $B\sim 1\,\mbox{T}$. Note that the energy separation between
the lowest Landau levels ($n=0,1$) and the $n=2$ level in the free bilayer model 
in a magnetic field is $\sqrt{2}\hbar \omega_c$,\cite{Falko} where the cyclotron energy is
$\hbar \omega_c = \hbar|eB|/mc \simeq 2.15 B\, [{\mbox T}] \, \mbox{meV}$ and the mass $m$ of quasiparticles
is $m \simeq 0.054 m_e$.
Even though it is of the same order of magnitude as the experimental gaps, the energy separation between
the Landau levels may be large enough to justify the use of the lowest Landau level (LLL) approximation in
the analysis of the gap equations.\cite{Barlas,Nandkishore1,GGM1,GGM2,Nandkishore2,Toke,Kharitonov1}

Recently, however, the experiments \cite{Freitag,Velasco} in suspended bilayer graphene with a higher mobility
($80,000$ to $100,000\,\mbox{cm}^2/\mbox{Vs}$) revealed much larger gaps, about $6.5\,\mbox{meV}$
at $B\simeq 1\,\mbox{T}$. It is natural that a higher mobility specimen produces much larger gaps compared
to those in Refs.~\onlinecite{Feldman,Weitz,Martin}. Formally, such  large gaps exceed the
energy of the 3rd Landau level in the free low-energy effective theory of bilayer graphene,\cite{Falko} i.e., in
a model with the Coulomb interaction ignored. Obviously, in this case the use of the LLL approximation in
the study of the generation of the gaps cannot be justified. In this study, therefore, we amend the analysis
of Refs.~\onlinecite{GGM1,GGM2} by properly taking {\em Landau level mixing} effects into account.\cite{footnote}

The inclusion of Landau level mixing is also instructive from the viewpoint of comparing
the quantum Hall ferromagnetism\cite{QHF} (QHF) and magnetic catalysis\cite{MC} (MC) scenarios describing the gap
generation in graphene (for example, see discussions in Refs.~\onlinecite{Yang,GGMS08,Goerbig,BYM}).
Technically, the difference between them comes from the choice of order parameters describing
the $SU(4)$ symmetry breaking. While the QHF and MC order parameters are inequivalent in general, they
are indistinguishable in the LLL approximation.\cite{GGM1,GGMS08,Goerbig} When the gaps are large and
Landau level mixing is essential, one cannot avoid/neglect any of these order parameters. Therefore, by solving
the corresponding gap equations and determining the ground state, one has an opportunity to shed some light
on the quantitative roles of the QHF and MC in the dynamics of graphene.

It is interesting to compare the importance of Landau level mixing in bilayer graphene with that in monolayer 
one.\cite{Semenoff,Kharitonov2,GGMS11.05}. The study in Ref.~\onlinecite{GGMS11.05} revealed several 
qualitative effects in the latter, including the ``running" of the gaps and the Fermi velocity with the
Landau level index $n$. A detailed analysis of the quasiparticle spectra in higher Landau levels ($n\geq 1$)
also reveals the role of QHF and MC order parameters. The same is expected in bilayer graphene. However, the
Coulomb interaction in bilayer should play a more profound role. This is the consequence
of a much smaller characteristic energy scale $\hbar \omega_{c}$ in bilayer graphene as compared to the Landau
energy scale $\varepsilon_{\ell}\simeq 26 \sqrt{B\, [{\mbox T}]}\, \mbox{meV}$ in monolayer one.

In this study, we make another essential improvement in the theoretical analysis\cite{GGM1,GGM2,GGJM}
by replacing the static (i.e., instantaneous) screening of the Coulomb interaction by the {\em dynamical} one.
As will be shown below, these two new ingredients lead to both larger gaps and an offset for the
critical value of the electric field $E_{\perp}^{\rm off}$, which separates the spin-polarized phase
and the layer-polarized one at zero magnetic field. The latter is observed in all experiments in
suspended bilayer graphene.\cite{Feldman,Weitz,Martin,Velasco}

Another interesting finding of our study of the dynamics with dynamically screened Coulomb
interaction in the $\nu = 0$ QH state is a Landau {\em level reordering}, caused by large gaps in
the lowest Landau levels. The result is a dramatic rearrangement of the quasiparticle spectrum: the
lowest conduction band and the highest valence one are now given by the $n= 2$ Landau level,
rather than by the LLL ($n=0,1$).

This paper is organized as follows. In Sec.~\ref{Model}, we introduce an effective low-energy
model of bilayer graphene. In this section, we also numerically calculate the frequency-dependent
polarization function. We find that the corresponding
function allows a simple fit given by the product of the static function and a simple form factor depending on
the energy and momentum. The polarization function is used in a coupled set of gap equations derived in
Sec.~\ref{Gap equation}. In Sec.~\ref{Numerical results}, we present our numerical results. A general
discussion of the main results and their comparison with experiment are presented in Sec.~\ref{Discussion}. 
Several appendixes at the end of the paper contain technical details and derivations used supplementing the
presentation in the main text.

\section{Model}
\label{Model}

We will utilize the same model as in Refs.~\onlinecite{GGM1,GGJM}. The free part of the effective low-energy
Hamiltonian of bilayer graphene reads\cite{Falko}
\begin{equation}
H_0 = - \frac{1}{2m}\sum_{\xi,s}\int
d^2\mathbf{r}\,\Psi_{\xi s}^{\dagger}(\mathbf{r})\left( \begin{array}{cc} 0 & (\pi^{\dagger})^2\\ \pi^2 & 0
\end{array} \right)\Psi_{\xi s}(\mathbf{r}), \label{free-Hamiltonian}
\end{equation}
where $\pi=\hat{p}_{x}+i\hat{p}_{y}$ and the canonical momentum
$\hat{\mathbf{p}} = -i\hbar\bm{\nabla}+ {e\mathbf{A}}/c$ includes the vector potential $\mathbf{A}$
corresponding to the external magnetic field
{$\mathbf{B}= (0_{\parallel}, B_{\perp})$, which is taken to be orthogonal to the bilayer planes,
$B \equiv |B_{\perp}|$.} The quasiparticle mass is $m= \gamma_1/2v_{F}^2\approx 0.054 m_e$, 
where $v_{F}\approx 8.0\times 10^{5}~\mbox{m/s}$ is the Fermi velocity, $\gamma_1 \approx 0.39~\mbox{eV}$,
and $m_e$ is the mass of the electron. The two component spinor field $\Psi_{\xi s}$ carries valley
($\xi=\pm$ for valley $K$ and $K^{\prime}$, respectively) and spin ($s = \downarrow, \uparrow$ for spin 
down and up, respectively) indices. We will use the standard  convention:\cite{Falko}
$\Psi_{+ s}^T=(\psi_{KA_1}, \psi_{KB_2})_{s}$ for valley $K$ and $\Psi_{- s}^T = (\psi_{K^{\prime} B_2},
\psi_{K^{\prime} A_1})_{s}$ for valley $K^{\prime}$. Indices $A_1$ and $B_2$ label the corresponding
$A$ and $B$ sublattices in the layers 1 (top) and 2 (bottom), respectively, which, according to Bernal 
$(A_2-B_1)$ stacking, are relevant for the low energy dynamics. Let us emphasize that the 
sublattice and layer degrees of freedom are not independent in this low-energy model: the sublattices 
A and B correspond to the layers $1$ and $2$, respectively.

\begin{widetext}
The Zeeman and Coulomb interactions plus a top-bottom gates voltage imbalance $\tilde{\Delta}_0$ in 
bilayer graphene are described as follows:
\begin{eqnarray}
H_{\rm int}&=& \sum_{\xi,s}\int d^2\mathbf{r}\,\Psi_{\xi s}^{\dagger}(\mathbf{r}) \left[
Z \sigma^3 +\tilde{\Delta}_0 \xi \tau^3-\frac{\tilde{\Delta}_0}{m\gamma_{1}}
\xi\left( 
\begin{array}{cc}\pi^{\dagger}\pi&0\\
0&-\pi\pi^{\dagger}\end{array}
\right)
\right]\Psi_{\xi s}(\mathbf{r})\, \nonumber \\
&+& \frac{1}{2}\int d^2\mathbf{r}d^2\mathbf{r}^{\prime}\,\Big\{
V(\mathbf{r}-\mathbf{r}^{\prime})\left[\rho_1(\mathbf{r})\rho_1(\mathbf{r}^{\prime})+
\rho_2(\mathbf{r})\rho_2(\mathbf{r}^{\prime})\right]  +
2V_{12}(\mathbf{r}-\mathbf{r}^{\prime})\rho_1(\mathbf{r})\rho_2(\mathbf{r}^{\prime})\Big\}\,,
\label{interaction1}
\end{eqnarray}
\end{widetext}
Here $Z \equiv \mu_{B}B=0.027\,\hbar\omega_c=0.059 B[T]\,\mbox{meV}$ is the Zeeman 
energy, $\sigma^3$ is the diagonal Pauli matrix in spin space, and $\tau^3$ is the diagonal Pauli 
matrix acting on the two components of the fields $\Psi_{+ s}$ and  $\Psi_{- s}$. 
Note that the presence of $\xi$ in the voltage imbalance term is related to the different order 
of the $A_1$ and $B_2$ components in $\Psi_{+ s}$ and $\Psi_{- s}$ [compare also with the 
layer projection operators in Eqs.~(\ref{density1}) and  (\ref{density}) below].
The voltage imbalance between 
the top and bottom gates is related to the electric field applied perpendicularly to the bilayer planes:  
$\tilde{\Delta}_0=e E_{\perp}d /2$.

The third term in the square brackets in the first line of Eq.~(\ref{interaction1})
leads to a small splitting of the Landau levels with orbital indices $n=0$ and $n=1$.\cite{Falko}
However, due to the factor $\gamma_{1}$ in the denominator, it
is suppressed in comparison to other terms in the interaction Hamiltonian (\ref{interaction1}). In fact,
the splitting that it produces is of order of $0.01\tilde{\Delta}_{0}B[T]$ that is substantially less
than the splittings due to the other two terms in the square brackets in the first line of Hamiltonian 
(\ref{interaction1}) for reasonable values of a magnetic field $B$ (for a recent discussion of this term, 
see Ref.~\onlinecite{GGJM}). As will be shown below, it is also much less than
the dynamical splitting between the $n=0$ and $n=1$ levels due to the Coulomb interaction.
Because of that, it will be omitted in the analysis below.

The Coulomb interaction
term $V(\mathbf{r})$ in $H_{\rm int}$ is the bare intralayer potential whose Fourier transform is given by
$\tilde{V}(p)={2\pi e^2}/(\kappa p)$, where $\kappa$ is the dielectric constant. The potential
$V_{12}(\mathbf{r})$ describes the interlayer electron interactions. Its Fourier transform is
$\tilde{V}_{12}(p)=2\pi e^2 \mbox{e}^{-pd}/(\kappa p)$, where $d=3.5\times 10^{-10} \,\mbox{m}$
is the distance between the layers. The two-dimensional charge densities in the two layers are
\begin{eqnarray}
\label{density1}
\rho_1(\mathbf{r})&=& \sum_{\xi,s}\Psi_{\xi s}^{\dagger}(\mathbf{r}){\cal P}_1(\xi)\Psi_{\xi s}(\mathbf{r})\,, \\
\rho_2(\mathbf{r})&=& \sum_{\xi,s}\Psi_{\xi s}^{\dagger}(\mathbf{r}){\cal P}_2(\xi)\Psi_{\xi s}(\mathbf{r})\,,
\label{density}
\end{eqnarray}
where ${\cal P}_1(\xi)=(1+ \xi\tau^3)/2$ and ${\cal P}_2(\xi)=(1 - \xi\tau^3)/2$
are projectors on the states in the layers 1 and 2,
respectively.
When the dynamical screening effects are taken into account, the potentials $V(\mathbf{r})$
and $V_{12}(\mathbf{r})$ are replaced by effective interactions $V_{\rm eff}(t,\mathbf{r})$ and
$V_{12\,{\rm eff}}(t,\mathbf{r})$ which are no longer instantaneous.

\begin{widetext}
The Schwinger--Dyson (gap) equation for the quasiparticle Green's function (propagator) in the Hartree-Fock
approximation reads:\cite{GGM2}
\begin{eqnarray}
G^{-1}(t-t^\prime;\bm{r},\bm{r}^\prime) &=& S^{-1}(t-t^\prime;\bm{r},\bm{r}^\prime)
- iG(t-t^\prime;\bm{r},\bm{r}^\prime) V_{\rm eff} (t-t^\prime;\bm{r}-\bm{r}^\prime) \nonumber  \\
&-& i\left[{\cal P}_1(\xi)G(t-t^\prime;\bm{r},\bm{r}^\prime) {\cal P}_2(\xi)
+{\cal P}_2(\xi)G(t-t^\prime;\bm{r},\bm{r}^\prime) {\cal P}_1(\xi)\right]
{V}_{\rm IL}(t-t^\prime;\bm{r-\bm{r}^\prime})   \nonumber \\
&-&\frac{i}{2}\left[{\cal P}_1(\xi)-{\cal P}_2(\xi)\right]\,\mbox{tr}\big\{\left[{\cal P}_1(\xi)-{\cal P}_2(\xi)\right]G(0;0)\big\}
\tilde{V}_{\rm IL}^{\rm bare}(0)  \delta(t-t^\prime)\delta(\bm{r}-\bm{r}^\prime),
\label{SD-equation}
\end{eqnarray}
\end{widetext}
where $V_{\rm IL}(t-t^\prime; \mathbf{r}-\mathbf{r}^\prime)
=V_{12\, {\rm eff}}(t-t^\prime; \mathbf{r}-\mathbf{r}^\prime)-V_{\rm eff}(t-t^\prime; \mathbf{r}-\mathbf{r}^\prime)$,
$S(t-t^\prime;\bm{r},\bm{r}^\prime)$ and $G(t-t^\prime;\bm{r},\bm{r}^\prime)$ are the free and full
Green's functions, respectively. These two Green's functions are described in Appendix~\ref{Quasiparticle propagator}.
In the presence of an external magnetic field, they are not translationally
invariant functions but can be written in the form of a product of a universal Schwinger phase
(which spoils the translational invariance) and translationally invariant functions.

The momentum space expressions for the interaction potentials in the momentum
space are\cite{GGM2,Potentials}
\begin{eqnarray}
\tilde{V}_{\rm eff}(\omega,p)  &=& \frac{2\pi e^2}{\kappa}\frac{1}{p+\frac{4\pi e^2}{\kappa}\Pi(\omega,p)},\\
\tilde{V}_{\rm IL}^{\rm bare}(0)  &=&-\frac{2\pi e^2 d}{\kappa}.
\end{eqnarray}
The explicit form of the interlayer potential $\tilde{V}_{\rm IL}(\omega,p)$ can be found in Appendix~A
of Ref.~\onlinecite{GGM2}. However, here we do not need it: due to the presence of
the projectors ${\cal P}_1(\xi)$ and ${\cal P}_2(\xi)$ in the second line in Eq.~(\ref{SD-equation}), 
the corresponding Fock term does not contribute to the final form of the gap
equation.  Note also that since bilayer has no net electric charge at the neutrality point,
we dropped all the Hartree terms proportional to $\mbox{tr}[G(0;0)]$ (i.e., the total density of charge carriers)
in the gap equation.\cite{footnote2}

\begin{figure*}
\begin{center}
\includegraphics[width=.42\textwidth]{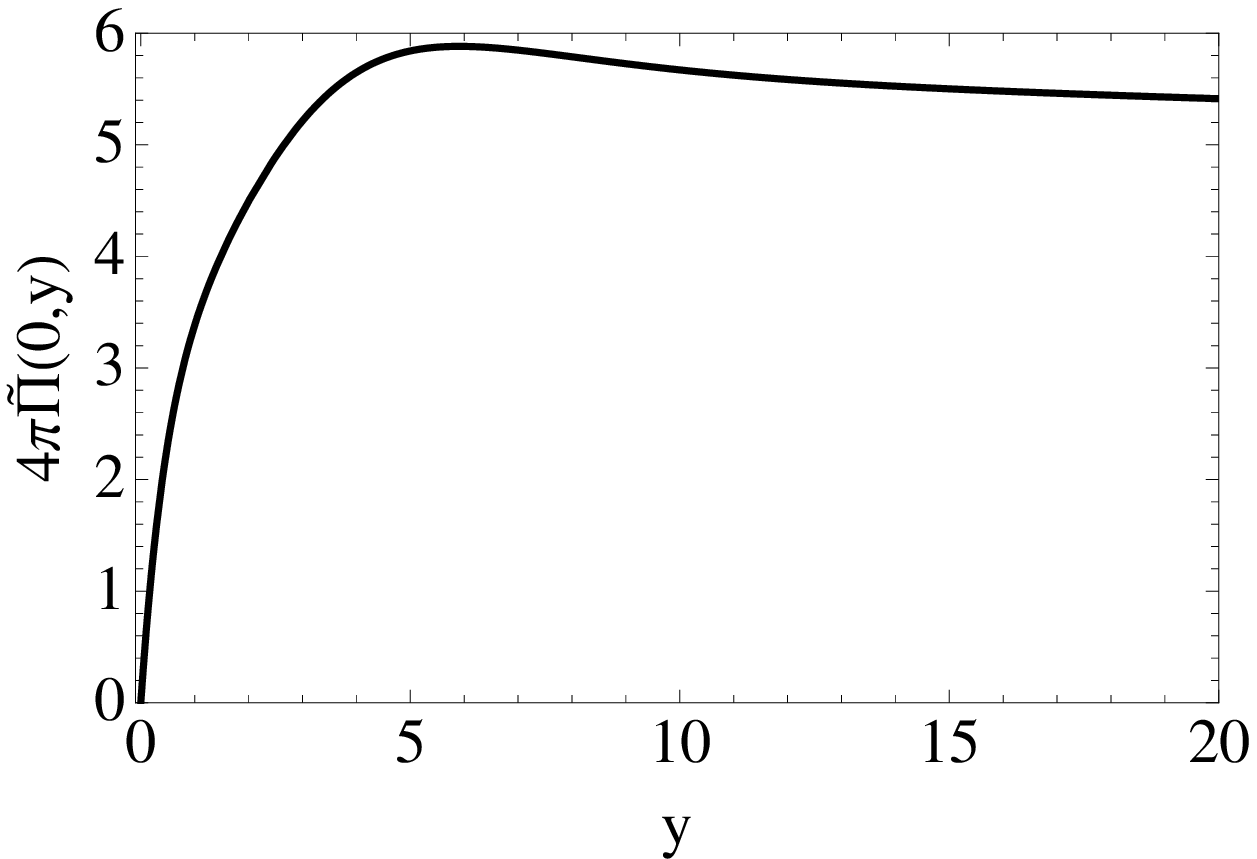}
\includegraphics[width=.55\textwidth]{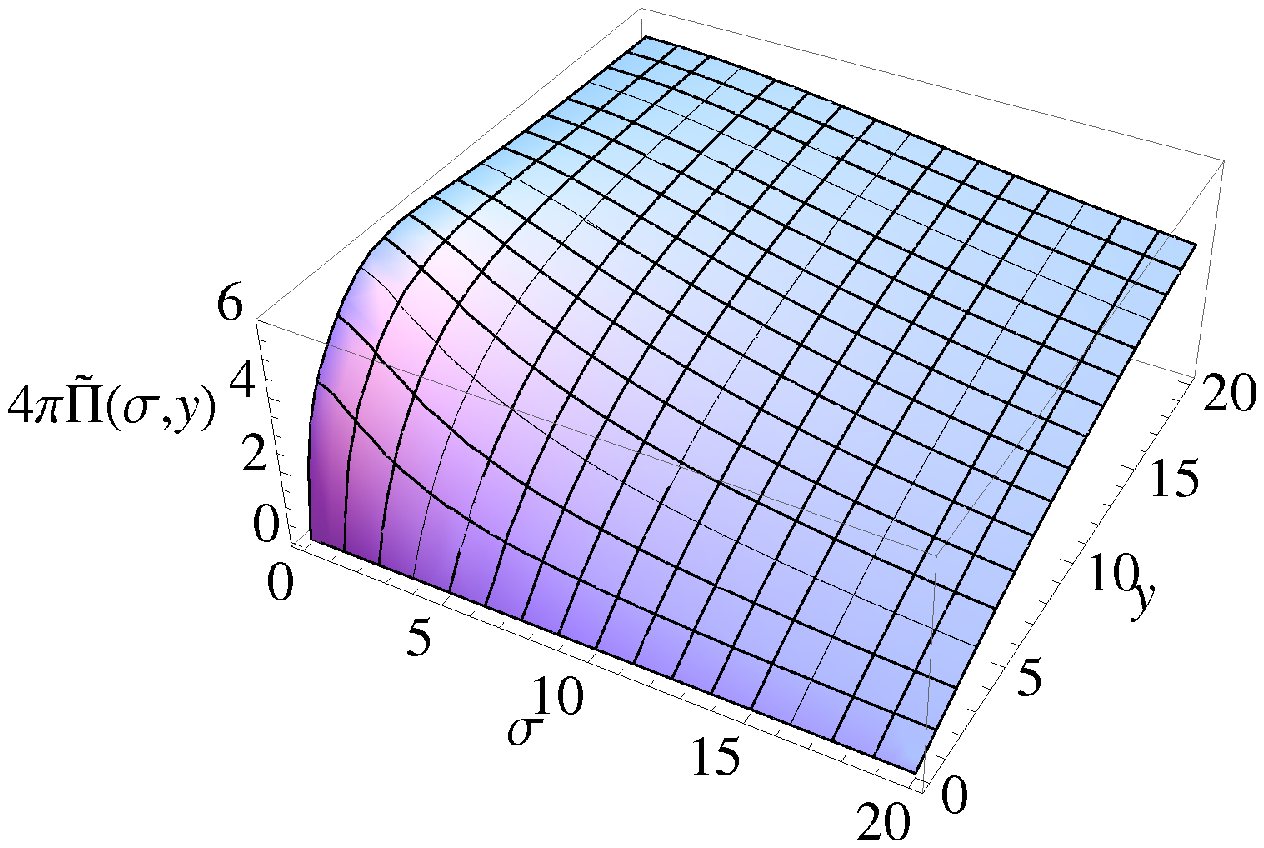}
\caption{(Color online) The static polarization function (left) and frequency-dependent
polarization function (right).}
\label{fig-pol-tensor}
\end{center}
\end{figure*}

The screened Coulomb interaction $\tilde{V}_{\rm eff}(\omega,p)$ depends on the dynamical polarization
function $\Pi(\omega,p)$. In the previous theoretical studies of the gap equation in bilayer graphene in a 
magnetic field, only a static polarization function was used.\cite{GGM1,GGM2,Nandkishore2,GGJM}
In the random phase approximation, the static polarization function in a magnetic field was calculated in 
Refs.~\onlinecite{GGM1,GGM2} and is replotted in the left panel in Fig.~\ref{fig-pol-tensor}.
The corresponding approximation for the interaction potential reads
\begin{eqnarray}
\tilde{V}_{\rm eff}(0;y)=\frac{2\pi \ell^2\hbar\omega_{c}}{\kappa\sqrt{x y}+4\pi \tilde{\Pi}(y)}\,,
\label{V-eff-static}
\end{eqnarray}
where  $y=(p \ell)^2/2$ is a dimensionless variable used instead of the wave vector $p$ and
$\Pi(0,p)=(m/\hbar^2)\tilde{\Pi}(y)$. By definition, $\ell \equiv \sqrt{\hbar c/|eB|}$ is the magnetic length 
and $x\equiv 2\hbar^4/(e^4m^2\ell^2) \approx 0.003 B~\mbox{[T]}$ is a dimensionless parameter 
which determines the value of $y$ below which the screening effects are negligible.

The use of static screening approximation greatly simplifies calculations. Our analysis below shows, however,
that this approximation significantly underestimates the strength of the Coulomb interaction in the Fock term.
In fact, we find that taking into account the effects of dynamical screening effectively leads to about three times
larger gaps compared to the case of the static screening.

\begin{widetext}
In the one loop approximation, the frequency-dependent polarization function in Euclidean space
(after the Wick rotation) is given by\cite{GGM2}
\begin{eqnarray}
\tilde{\Pi}(\sigma,y)&=& \frac{1}{\pi}\sum_{n=2}^{\infty}  \frac{{\cal M}_{n}}{{\cal M}_{n}^2+\sigma^2}
\left[{\cal L}_{0,n}(y)+{\cal L}_{1,n}(y)\right]\nonumber\\
&+&\frac{1}{2\pi}\sum_{n,m=2}^{\infty} \frac{{\cal M}_{n}+{\cal M}_{m}}{({\cal M}_{n}+{\cal M}_{m})^2+\sigma^2}
\left({\cal L}_{n,m}(y)+{\cal L}_{n-2,m-2}(y)-\frac{2}{{\cal M}_{n} {\cal M}_{m}}{\cal L}^{(2)}_{n-2,m-2}(y)\right),
\label{dynamicPolarization}
\end{eqnarray}
where the dimensionless parameters ${\cal M}_{n}=\sqrt{n(n-1)}$, $\sigma=\omega/\omega_{c}$, and
the following functions were introduced:
\begin{eqnarray}
{\cal L}^{(\alpha)}_{n,m}&=&\frac{1}{2\pi l^{2}}\int d^{2}r\,e^{-i\mathbf{p}\mathbf{r}}
\left(\frac{\mathbf{r}^{2}}{2l^{2}}\right)^{\alpha}e^{-\mathbf{r}^{2}/2l^{2}}
L_{n}^{\alpha}\left(\frac{\mathbf{r}^{2}}{2l^{2}}\right)L_{m}^{\alpha}\left(\frac{\mathbf{r}^{2}}{2l^{2}}\right)
=\int\limits_{0}^{\infty}dtt^{\alpha}e^{-t}J_{0}(2\sqrt{y t})L_{n}^{\alpha}(y)L_{m}^{\alpha}(y)\nonumber\\
&=&(-1)^{n+m}\frac{(m+\alpha)!}{m!}e^{-y}L_{m+\alpha}^{n-m}(y)L_{n}^{m-n}(y).
\end{eqnarray}
\end{widetext}
Here  $L_{n}^{\alpha}\left(y\right)$ are the generalized Laguerre polynomials and $J_{0}(x)$
is the Bessel function {(${\cal L}_{n,m} \equiv {\cal L}^{(0)}_{n,m}$).}
(For calculation of such integrals, see Appendix A in Ref.~\onlinecite{GGM2}.)
In the case of $\alpha=0,2$, in particular, one obtains
\begin{equation}
{\cal L}_{n,m}(y) = (-1)^{n+m}e^{-y}L_{m}^{n-m}\left(y\right) L_{n}^{m-n}\left(y\right),
\label{L_nm}
\end{equation}
\begin{equation}
{\cal L}^{(2)}_{n,m}(y) = (-1)^{n+m}e^{-y}(m+1)(m+2)L_{m+2}^{n-m}\left(y\right) L_{n}^{m-n}\left(y\right).
\end{equation}
The above expression for the dynamical polarization function can be calculated numerically with a relative
ease by employing a simple trick. First, we subtract the static part from Eq.~(\ref{dynamicPolarization})
to obtain an expression for the difference $\tilde{\Pi}(\sigma,y)-\tilde{\Pi}(0,y)$. Such a difference is given
in terms of a rather quickly convergent series and, therefore, can be tabulated as a function of two variables,
$\sigma$ and $y$, without much effort. After adding the static part, which is a function of only one variable
and which was calculated earlier,\cite{GGM1,GGM2} we finally obtain the polarization function $\tilde{\Pi}(\sigma,y)$.
It is plotted in the right panel in Fig.~\ref{fig-pol-tensor}. As expected, the dynamical effects reduce screening
at nonzero $\omega$.

We find that, in a wide range of momenta and frequencies, $0\leq (p\ell)^2/2\leq 10$ and
$0\leq \omega/\omega_{c}\leq 10$, the dynamical polarization function can be well approximated
(to within a few percent) by the following fit:
\begin{equation}
\tilde{\Pi}_{\rm fit}(\sigma,y) = \tilde{\Pi}(0,y) \frac{1+b_{1}y }{\sqrt{1+b_{2}y+b_{3}y^{2}
+b_{4}\sigma^{2} }}\,,
\label{polar-tensor-model}
\end{equation}
where the fit parameters are $b_{1} = 0.608$,  $b_{2} =1.572$,  $b_{3} = 0.357$, and $b_{4} = 0.868$.
Below this fit is used in our analysis of the dynamics to model the interaction potential with dynamical
screening, i.e.,
\begin{eqnarray}
\tilde{V}_{\rm eff}(\omega;y)=\frac{2\pi \ell^2\hbar\omega_{c}}{\kappa\sqrt{x y}+4\pi \tilde{\Pi}_{\rm fit}(\sigma,y)}\,,
\label{V-eff-dynamical}
\end{eqnarray}
which will replace the approximation with static screening in Eq.~(\ref{V-eff-static}).

\section{Gap equation}
\label{Gap equation}

As discussed in the Introduction, here we consider a rather general ansatz for the full Green's function
that includes the effects of both QHF and MC 
order parameters. While the former describe chemical potentials connected with conserved
charges, the latter describe Dirac like masses. 

As discussed at length in Ref.~\onlinecite{GGM2}, if both the Zeeman and $\tilde{\Delta}_0$ 
terms are ignored, the Hamiltonian $H = H_0 + H_{\rm int}$, with $H_0$ and $H_{\rm int}$ in 
Eqs.~(\ref{free-Hamiltonian}) and (\ref{interaction1}), possesses the symmetry 
$G = U^{(K)}(2)_s \times U^{(K^{\prime})}(2)_s \times Z_{2V}^{(\uparrow)}\times Z_{2V}^{(\downarrow)}$, where 
$U^{(V)}(2)_s$ defines the $U(2)$ spin transformations in a fixed valley $V = K, K^{\prime}$, 
and $Z_{2V}^{(s)}$ describes the valley transformation $\xi \to -\xi$ for a fixed spin 
$(s = \downarrow, \uparrow)$. The Zeeman interaction lowers this symmetry down to 
$G_2 \equiv U^{(K)}(1)_{\uparrow} \times U^{(K)}(1)_{\downarrow} \times U^{(K^{\prime})}(1)_{\uparrow} \times 
U^{(K^{\prime})}(1)_{\downarrow} \times Z_{2V}^{(\uparrow)}\times Z_{2V}^{(\downarrow)}$, where $U^{(V)}(1)_{s}$ is the
$U(1)$ transformation for fixed values of both valley and spin. Including the $\tilde{\Delta}_0$ 
term lowers the $G_2$ symmetry further down to the $\bar{G}_2 \equiv U^{(K)}(1)_{\uparrow}\times 
U^{(K)}(1)_{\downarrow} \times U^{(K^{\prime})}(1)_{\uparrow} \times U^{(K^{\prime})}(1)_{\downarrow}$.

The dynamics in the integer QH effect in bilayer graphene is intimately
connected with dynamical breakdown of the $G$ and $G_2$ symmetries.
Two sets of the order parameters describing their breakdown were considered in
Refs.~\onlinecite{GGM1,GGM2}. The first set consists of the quantum Hall ferromagnetism (QHF)
order parameters:\cite{QHF}
\begin{eqnarray}
\label{mu}
\mu_3: &\quad &
\sum_{\xi,s} \langle\Psi_{\xi,s}^\dagger s \Psi_{\xi,s}\rangle  \,,\\
\label{mu-tilde}
\tilde{\mu}_{s}: &\quad &
\sum_{\xi}  \langle\Psi^\dagger_{\xi,s}\xi\Psi_{\xi,s}\rangle \,.
\end{eqnarray}
While the order parameter (\ref{mu}) is a conventional ferromagnetic one,
the order parameter (\ref{mu-tilde}) determines the charge-density
imbalance between the two valleys. The corresponding chemical potentials
are $\mu_3$ and $\tilde{\mu}_s$, respectively.

The second set consists of the magnetic catalysis (MC) order parameters,\cite{MC}
i.e., the Dirac $\tilde{\Delta}_s$ and Haldane $\Delta_s$ mass terms:
\begin{eqnarray}
\label{delta}
\Delta_s: &\quad &
\sum_{\xi}  
\langle \Psi^\dagger_{\xi,s}\tau^3\Psi_{\xi,s}\rangle  \, ,\\
\label{delta-tilde}
\tilde{\Delta}_{s}: &\quad &
\sum_{\xi}  
\langle\Psi^\dagger_{\xi,s}\xi\tau^3\Psi_{\xi,s}\rangle  \,.
\end{eqnarray}
The order parameter (\ref{delta}) describes a charge density wave in both the $K$ and $K^{\prime}$ 
valleys. This order parameter preserves the $G_2$ symmetry.\cite{Hald} 
On the other hand, the order parameter (\ref{delta-tilde}), connected with the conventional Dirac mass 
$\tilde{\Delta}_s$, determines the charge-density
imbalance between the two layers.\cite{Falko} The structure of this mass term coincides with that   
of the bare voltage imbalance term between the top and bottom gates introduced in Hamiltonian
(\ref{interaction1}) and, therefore, can be considered as a dynamical counterpart of the latter.
Like the QHF order parameter
(\ref{mu-tilde}), this mass term completely breaks the $Z_{2V}^{(s)}$
symmetry.

It is important that
in both monolayer and bilayer graphene, these two sets of the order parameters necessarily
coexist\cite{GGM1,GGM2,GGMS08} and are produced even at the weakest repulsive interactions between
electrons (magnetic catalysis\cite{catal,Khvesh,graphite}). The essence of this phenomenon
is an effective reduction by two units of the spatial dimension in the
electron-hole pairing in the LLL with energy $E=0$.

Beside these parameters, the Green's function includes also the chemical potential
$\mu$ related to the charge density $\sum_{\xi,s} \langle\Psi_{\xi,s}^\dagger\Psi_{\xi,s}\rangle$. Because of that, it will be
convenient to consider their combinations $\mu_\uparrow = \mu + \mu_3$ and $\mu_\downarrow = \mu - \mu_3$.      
Let us emphasize that all the parameters ($\mu_s$, $\tilde{\mu}_s$, $\Delta_s$, and $\tilde{\Delta}_s$) 
should be viewed as
operators with well defined values only when projected onto quasiparticle quantum states, in which case
they become functions of the Landau level index $n$: $\mu_{ns}$, $\tilde{\mu}_{ns}$, $\Delta_{ns}$,
and $\tilde{\Delta}_{ns}$. The derivations and explicit expressions of the full/free quasiparticle
Green's functions are presented in Appendix~\ref{Quasiparticle propagator}.

Concerning the full Green's function, it is interesting to note that the order parameters in the lowest
Landau level enter the quasiparticle energies only through two independent combinations:\cite{GGM2}
$E^{L}_{\xi ns}=-\left(\mu_{ns}+\Delta_{ns}\right)+\xi (\tilde{\mu}_{ns}-\tilde{\Delta}_{ns})$, where $n=0,1$.
This is the consequence of the spinor structure in LLL wave functions, which have only half of the
components nonvanishing (the valley and layer degrees of freedom are not independent
in the LLL). Formally, this is also the reason why the roles of QHF and MC order parameters are
indistinguishable in the LLL approximation. Without the loss of generality, we will assume in the
following that only {$\Delta_{ns}$} and $\tilde{\Delta}_{ns}$ are nontrivial parameters in the lowest
Landau level ($n=0,1$).

\begin{widetext}
According to Eqs.~(\ref{A9}) and (\ref{A15}) in Appendix~\ref{Quasiparticle propagator},
the Green's function and its inverse contain the same overall Schwinger phase, which
describes the breakdown of the conventional translation invariance in a magnetic field.
After substituting these functions into the gap equation (\ref{SD-equation}), omitting the phase on both sides,
and performing a Fourier transform with respect to the time variable, we arrive at the following equation
for the translationally invariant part of the Green's function:
\begin{eqnarray}
\tilde{G}^{-1}(\omega;\bm{r}) &=& \tilde{S}^{-1}(\omega;\bm{r})
-i\int \frac{d\omega^\prime}{2\pi}\int \frac{p dp}{2\pi} J_{0}(pr)\,
\tilde{G}(\omega^\prime;\bm{r})\, \tilde{V}_{\rm eff} (\omega^\prime-\omega;p)\nonumber\\
&-&\frac{i}{2}\int \frac{d\omega^\prime}{2\pi}\int \frac{pdp}{2\pi} J_{0}(pr)\left[
\tilde{G}(\omega^\prime;\bm{r}) - \tau_{3}\tilde{G}(\omega^\prime;\bm{r}) \tau_{3}
\right] \tilde{V}_{\rm IL} (\omega^\prime-\omega;p)\nonumber\\
&-&\frac{i}{2}\int \frac{d\omega^\prime}{2\pi}\left[{\cal P}_1(\xi)-{\cal P}_2(\xi)\right]\,
\mbox{tr}\left\{\left[{\cal P}_1(\xi)-{\cal P}_2(\xi)\right]\tilde{G}(\omega^\prime;0)\right\}
\tilde{V}^{\rm bare}_{\rm IL}(0) \delta(\bm{r})\,.
\label{SD-translation-invariant}
\end{eqnarray}
After substituting the explicit form of the translationally invariant functions
given in Eqs.~(\ref{A10}) and (\ref{A16}), we calculate the traces and obtain equations
for the order parameters $\mu$, $\tilde{\mu}$, $\Delta$, and $\tilde{\Delta}$.

Furthermore, assuming these order
parameters are energy independent,  we can analytically integrate over the energy
and momentum on the right hand side of the gap equation (\ref{SD-translation-invariant}).
Finally, after projecting out onto different orbitals (associated with different Laguerre
polynomials), we arrive at a final set of algebraic equations for the order parameters
in all Landau levels:
\begin{eqnarray}
-E_{\xi 0 s}^{L} -\bar{\mu}_{s}-\xi \tilde{\Delta}_{0}
&=&\frac{\hbar\omega_{c}}{2}  \left[
- I_{0,0}(E_{\xi 0 s}^{L} ) \, \mbox{sign}(E_{\xi 0 s}^{L} ) - I_{1,0}(E_{\xi 1 s}^{L} ) \,
\mbox{sign}(E_{\xi 1 s}^{L} ) \right] \nonumber\\
&+&\frac{\hbar\omega_{c}}{4} \sum_{n^{\prime}=2}^{n_{\rm max}} I_{n^{\prime},0}(M_{\xi n^{\prime}s}
+\mu_{\xi n^{\prime}s}) \, \mbox{sign}(M_{\xi n^{\prime}s}+ \mu_{\xi n^{\prime}s})
\left(1+ \frac{ \Delta_{\xi n^{\prime}s} }{ M_{\xi n^{\prime}s}} \right)
\nonumber\\
&-&\frac{\hbar\omega_{c}}{4}\sum_{n^{\prime}=2}^{n_{\rm max}} I_{n^{\prime},0} (M_{\xi n^{\prime}s}
- \mu_{\xi n^{\prime}s}) \, \mbox{sign}(M_{\xi n^{\prime}s}- \mu_{\xi n^{\prime}s})
\left(1- \frac{ \Delta_{\xi n^{\prime}s} }{ M_{\xi n^{\prime}s}} \right)
\nonumber\\
&+& \frac{\alpha_{\rm IL} \xi  \hbar\omega_c}{2} \sum_{\xi^{\prime},s^{\prime}}\xi^{\prime}
\left[ \frac{1}{2} \mbox{sign}(E_{\xi^{\prime} 0 s^{\prime}}^{L})
+\frac{1}{2} \mbox{sign}(E_{\xi^{\prime} 1 s^{\prime}}^{L})
-\sum_{n^{\prime}=2}^{n_{\rm max}}
\frac{\Delta_{\xi^{\prime} n^{\prime}s^{\prime}}}{M_{\xi^{\prime} n^{\prime}s^{\prime}}}
\theta(M_{\xi^{\prime} n^{\prime}s^{\prime}}^{2}-\mu_{\xi^{\prime} n^{\prime}s^{\prime}}^2) \right],
\label{dyn-gap-eq-n=0}
\end{eqnarray}
\begin{eqnarray}
-E_{\xi 1 s}^{L} -\bar{\mu}_{s}-\xi \tilde{\Delta}_{0}
&=&\frac{\hbar\omega_{c}}{2}  \left[
-I_{0,1}(E_{\xi 0 s}^{L} )\mbox{sign}(E_{\xi 0 s}^{L} ) -I_{1,1}(E_{\xi 1 s}^{L} )
\mbox{sign}(E_{\xi 1 s}^{L} ) \right]\nonumber\\
&+&\frac{\hbar\omega_{c}}{4}\sum_{n^{\prime}=2}^{n_{\rm max}} I_{n^{\prime},1}(M_{\xi n^{\prime}s}
+\mu_{\xi n^{\prime}s}) \, \mbox{sign}(M_{\xi n^{\prime}s}+ \mu_{\xi n^{\prime}s})
\left(1+ \frac{ \Delta_{\xi n^{\prime}s} }{ M_{\xi n^{\prime}s}} \right)
\nonumber\\
&-&\frac{\hbar\omega_{c}}{4}\sum_{n^{\prime}=2}^{n_{\rm max}} I_{n^{\prime},1} (M_{\xi n^{\prime}s}
-\mu_{\xi n^{\prime}s}) \, \mbox{sign}(M_{\xi n^{\prime}s}- \mu_{\xi n^{\prime}s})
\left(1- \frac{ \Delta_{\xi n^{\prime}s} }{ M_{\xi n^{\prime}s}} \right)
\nonumber\\
& +& \frac{\alpha_{\rm IL} \xi  \hbar\omega_c}{2} \sum_{\xi^{\prime},s^{\prime}}\xi^{\prime}
\left[\frac{1}{2} \mbox{sign}(E_{\xi^{\prime} 0 s^{\prime}}^{L})
+\frac{1}{2} \mbox{sign}(E_{\xi^{\prime} 1 s^{\prime}}^{L})
-\sum_{n^{\prime}=2}^{n_{\rm max}}
\frac{\Delta_{\xi^{\prime} n^{\prime}s^{\prime}}}{M_{\xi^{\prime} n^{\prime}s^{\prime}}}
\theta(M_{\xi^{\prime} n^{\prime}s^{\prime}}^{2}-\mu_{\xi^{\prime} n^{\prime}s^{\prime}}^2) \right]  ,
\label{dyn-gap-eq-n=1}
\end{eqnarray}
for $n=0$ and $n=1$, respectively, and a pair of equations,
\begin{eqnarray}
\hspace*{-5mm}
\mu_{\xi ns}-\bar{\mu}_{s}+\Delta_{\xi ns}-\xi \tilde{\Delta}_{0}&=&
\frac{\hbar\omega_{c}}{2}  \left[
-I_{0,n}(E_{\xi 0 s}^{L} )\mbox{sign}(E_{\xi 0 s}^{L} ) -I_{1,n}(E_{\xi 1 s}^{L} ) \mbox{sign}(E_{\xi 1 s}^{L} ) \right]
\nonumber\\
&+&\frac{\hbar\omega_{c}}{4}\sum_{n^{\prime}=2}^{n_{\rm max}} I_{n^{\prime},n}(M_{\xi ns}
+\mu_{\xi n^{\prime}s}) \, \mbox{sign}(M_{\xi n^{\prime}s}+ \mu_{\xi n^{\prime}s})
\left(1+ \frac{ \Delta_{\xi n^{\prime}s} }{ M_{\xi n^{\prime}s}} \right)
\nonumber\\
&-&\frac{\hbar\omega_{c}}{4}\sum_{n^{\prime}=2}^{n_{\rm max}} I_{n^{\prime},n} (M_{\xi n^{\prime}s}
-\mu_{\xi n^{\prime}s}) \, \mbox{sign}(M_{\xi n^{\prime}s}- \mu_{\xi n^{\prime}s})
\left(1- \frac{ \Delta_{\xi n^{\prime}s} }{ M_{\xi n^{\prime}s}} \right)
\nonumber\\
&+&  \frac{\alpha_{\rm IL} \xi  \hbar\omega_c}{2} \sum_{\xi^{\prime},s^{\prime}}\xi^{\prime}
 \left[\frac{1}{2} \mbox{sign}(E_{\xi^{\prime} 0 s^{\prime}}^{L})
+\frac{1}{2} \mbox{sign}(E_{\xi^{\prime} 1 s^{\prime}}^{L})
-\sum_{n^{\prime}=2}^{n_{\rm max}}
\frac{\Delta_{\xi^{\prime} n^{\prime}s^{\prime}}}{M_{\xi^{\prime} n^{\prime}s^{\prime}}}
\theta(M_{\xi^{\prime} n^{\prime}s^{\prime}}^{2}-\mu_{\xi^{\prime} n^{\prime}s^{\prime}}^2)\right],
\label{dyn-gap-eq-n>1A}
\\
\hspace*{-5mm}
\mu_{\xi ns}-\bar{\mu}_{s}-\Delta_{\xi ns}+\xi \tilde{\Delta}_{0}
&=&\frac{\hbar\omega_{c}}{4}\sum_{n^{\prime}=2}^{n_{\rm max}} I_{n^{\prime}-2,n-2}(M_{\xi n^{\prime}s}
+\mu_{\xi n^{\prime}s}) \, \mbox{sign}(M_{\xi n^{\prime}s}+ \mu_{\xi n^{\prime}s})
\left(1- \frac{ \Delta_{\xi n^{\prime}s} }{ M_{\xi n^{\prime}s}} \right)
\nonumber\\
&-&\frac{\hbar\omega_{c}}{4}\sum_{n^{\prime}=2}^{n_{\rm max}} I_{n^{\prime}-2,n-2} (M_{\xi n^{\prime}s}
-\mu_{\xi n^{\prime}s}) \, \mbox{sign}(M_{\xi n^{\prime}s}- \mu_{\xi n^{\prime}s})
\left(1+ \frac{ \Delta_{\xi n^{\prime}s} }{ M_{\xi n^{\prime}s}} \right)  \nonumber\\
&-& \frac{\alpha_{\rm IL} \xi  \hbar\omega_c}{2} \sum_{\xi^{\prime},s^{\prime}}\xi^{\prime}
\left[\frac{1}{2} \mbox{sign}(E_{\xi^{\prime} 0 s^{\prime}}^{L})
+\frac{1}{2} \mbox{sign}(E_{\xi^{\prime} 1 s^{\prime}}^{L})
-\sum_{n^{\prime}=2}^{n_{\rm max}}
\frac{\Delta_{\xi^{\prime} n^{\prime}s^{\prime}}}{M_{\xi^{\prime} n^{\prime}s^{\prime}}}
\theta(M_{\xi^{\prime} n^{\prime}s^{\prime}}^{2}-\mu_{\xi^{\prime} n^{\prime}s^{\prime}}^2)
\right] ,
\label{dyn-gap-eq-n>1B}
\end{eqnarray}
for each $n\geq 2$. Here, $\bar{\mu}_{s}=\mu_{0}-s Z$ is the effective chemical potential that
includes the shift due to the Zeeman energy ($s=\pm1$ corresponds to up and down spins,
respectively), $\mu_{0}$ is the chemical potential itself, and
$\alpha_{\rm IL}=e^2\gamma_1 d/(2\hbar^2v_{F}^2\kappa)\approx 0.354/\kappa$ is a dimensionless
interlayer coupling constant. 
\end{widetext}

The other notations are 
\begin{eqnarray}
\mu_{\xi ns}&=& \mu_{ns}-\xi\tilde{\mu}_{ns},\\
\Delta_{\xi ns}&=& \xi \tilde{\Delta}_{ns}+\Delta_{ns},\\
M_{\xi ns}&=&\sqrt{\Delta_{\xi ns}^2+ (\hbar\omega_{c})^2n(n-1)},\quad \mbox{for} \quad n\geq 2,\\
\label{LLLn}
E_{\xi n s}^{L} &=& - \Delta_{ns}  - \xi \tilde{\Delta}_{ns},\quad \mbox{for} \quad n=0,1.
\end{eqnarray}
It is shown in Appendix \ref{Quasiparticle propagator} that while $E_{\xi n s}^{L} $ in Eq.~(\ref{LLLn})
are the quasiparticle energies in the two lowest Landau levels, the quasiparticle energies
in higher Landau levels ($n\geq 2$) are given by the following expression:
\begin{eqnarray}
E_{\xi n s} &=& -\mu_{\xi ns}\pm M_{\xi ns},\quad \mbox{for} \quad n\geq 2.
\label{2n}
\end{eqnarray}
The set of gap equations (\ref{dyn-gap-eq-n=0}) -- (\ref{dyn-gap-eq-n>1B}) determines 
the order parameters $\mu$, $\tilde{\mu}$, $\Delta$, and $\tilde{\Delta}$ as functions of the Landau level
index $n$ and the external parameters: the chemical potential $\mu_{0}$, the magnetic field $B$,
and the applied electric field $E_{\perp}$, which is expressed through the 
top-bottom gate imbalance $\tilde{\Delta}_{0}$, $E_{\perp}=2 \tilde{\Delta}_0/(ed)$.\cite{Falko}
The  energy-dependent coefficient functions $I_{n^{\prime},n}(E)$ in these equations
are given in Eq.~(\ref{Innprime}) in Appendix~\ref{Tables}.
Each of the dimensionless coefficient functions $I_{n^{\prime},n}(E)$ can be calculated numerically with a
relative ease. However, when repeatedly solving a complete set of gap equations with many Landau levels, it is
greatly beneficial to use either a tabulated set of these functions or approximate analytical expressions.
It appears that the numerical results for functions $I_{n^{\prime},n}(E)$ can be well approximated by
the following functional dependence on the energy:
\begin{equation}
I_{n^{\prime},n}(E) = \frac{a_{n^{\prime},n}+ b_{n^{\prime},n} \frac{|E|}{\hbar\omega_c} + c_{n^{\prime},n}
\left(\frac{E}{\hbar\omega_c} \right)^2}{1+d_{n^{\prime},n} \frac{|E|}{\hbar\omega_c} } ,
\label{Innprime-Pade}
\end{equation}
which is usually valid in a wide range of energies ($0<E\lesssim 40 \hbar\omega_c$). In the case of
$B=2\,\mbox{T}$ and $\kappa=2$, for example, the corresponding coefficients for $0\leq n^{\prime},n\leq 10$
are summarized in Table~\ref{tab-a-b-c-d} in Appendix~\ref{Tables}. We show only the upper triangular part of
the corresponding tables because all coefficients are symmetric with respect to the exchange $n^{\prime}$ and $n$.

The approximation with static screening can be easily obtained from the more general dynamical one in
Eq.~(\ref{Innprime}) by substituting $E=0$ and taking $b_i=0$ ($i=1,4$).
The corresponding numerical calculation becomes very easy and one can tabulate the static coefficients
$I_{n^{\prime},n}$ on the fly.

Before concluding this section, let us also briefly discuss the truncation of a formally infinite set of the
gap equations. From a physics viewpoint, it is clear that the low-energy effective model is valid only in
a finite range of energies (up to about energy $\Lambda\approx \gamma_{1}/4\simeq 1000\,\mbox{K}$).\cite{Falko}
Therefore, the sum over Landau levels in Eqs.~(\ref{dyn-gap-eq-n=0}) -- (\ref{dyn-gap-eq-n>1B})
should be truncated at $n_{\rm max} \simeq k_B\Lambda/(\hbar\omega_c) \simeq 40/(B\,\mbox{[T]})$, i.e., 
$n_{\rm max}$ is different for different values of $B$ and it decreases with increasing $B$. The use of such 
a prescription implies that the model at hand is appropriate only for magnetic fields $B \lesssim 40\,\mbox{T}$.
In order to consider stronger magnetic fields, $B \gtrsim 40\,\mbox{T}$, or include Landau levels beyond 
$n_{\rm max}$, one should utilize the  microscopic four-band model.\cite{Falko} This is beyond the scope of 
the present work. In the opposite case
of weak magnetic fields, $B \lesssim 1\,\mbox{T}$, the model becomes expensive numerically and should be 
treated differently. In particular, the limit $B \to 0$ (no magnetic field) can be described only if an infinite number 
of Landau levels are taken into account, which would require a different approach to the problem.

\section{Numerical results}
\label{Numerical results}

In this study we concentrate on the ground states at the neutrality point. For the purposes of our analysis,
it is sufficient to fix the value of the chemical potential $\mu_0=0$. We will see that there are two main
solutions to the gap equation: the spin- and layer-polarized ones.\cite{footnote3} In order to determine which of them
corresponds to a true ground state of bilayer at given values of gates bias and magnetic field, we must
compare their free energy densities. The corresponding expression for the free energy density is derived in
Appendix~\ref{Free energy}.

{\subsection{Landau level mixing in static screening approximation}}
\label{StaticApproximation}

Before we proceed to a more complicated analysis of the gap equations with dynamical
screening in Subsec.~\ref{DynamicalApproximation}, let us first present our main results
for the static screening case. Such an approximation is instructive for getting a better
understanding of the Landau level mixing effects.

\begin{figure*}
\begin{center}
\includegraphics[width=.8\textwidth]{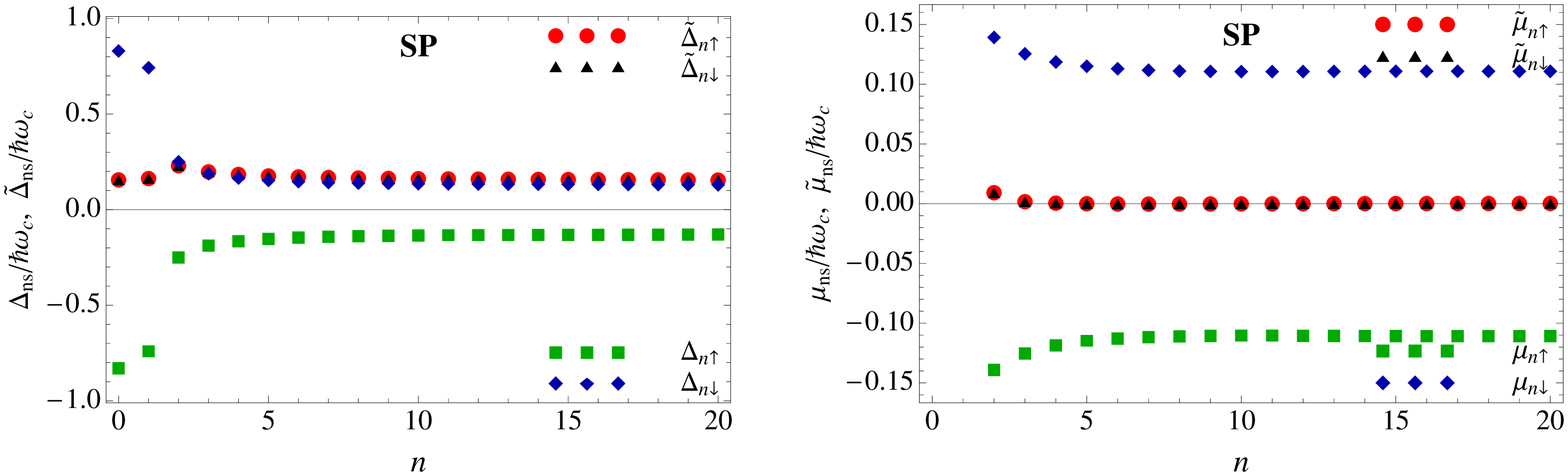}\\
\includegraphics[width=.8\textwidth]{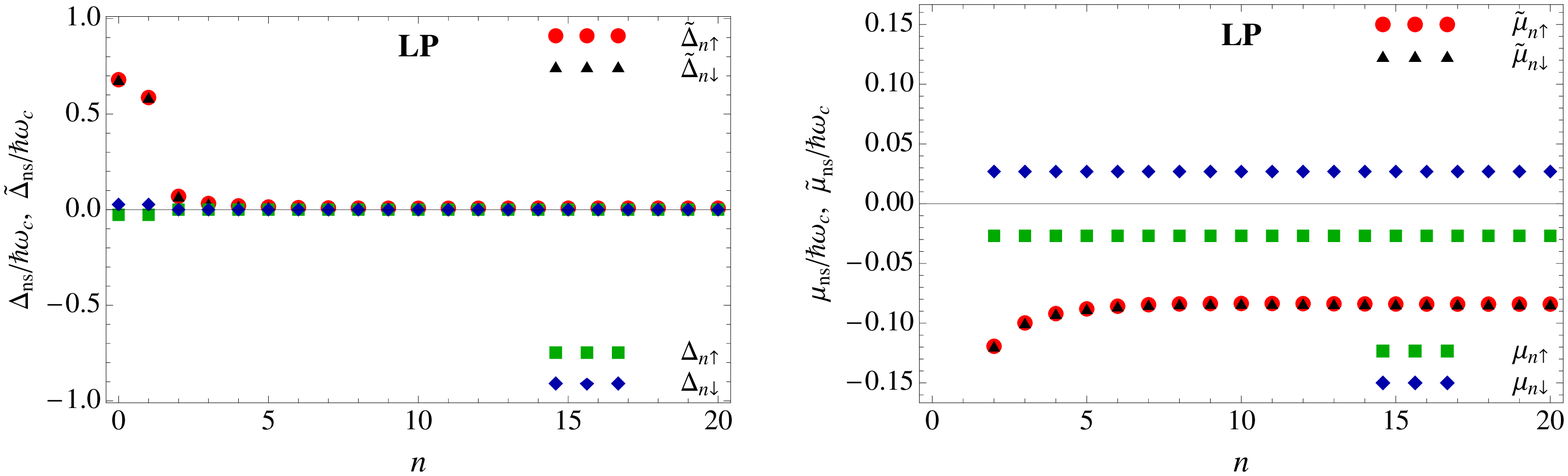}
\caption{(Color online) The dependence of the dynamical parameters on the Landau level index
$n$ for the SP (top) and LP (bottom) solutions at a fixed value of 
$\tilde{\Delta}_{0}\approx 0.3 \hbar\omega_c$ in the approximation with static screening. 
The magnetic field is $B= 2\,\mbox{T}$.}
\label{pars-vs-n-static}
\end{center}
\end{figure*}

\begin{figure*}
\begin{center}
\includegraphics[width=.32\textwidth]{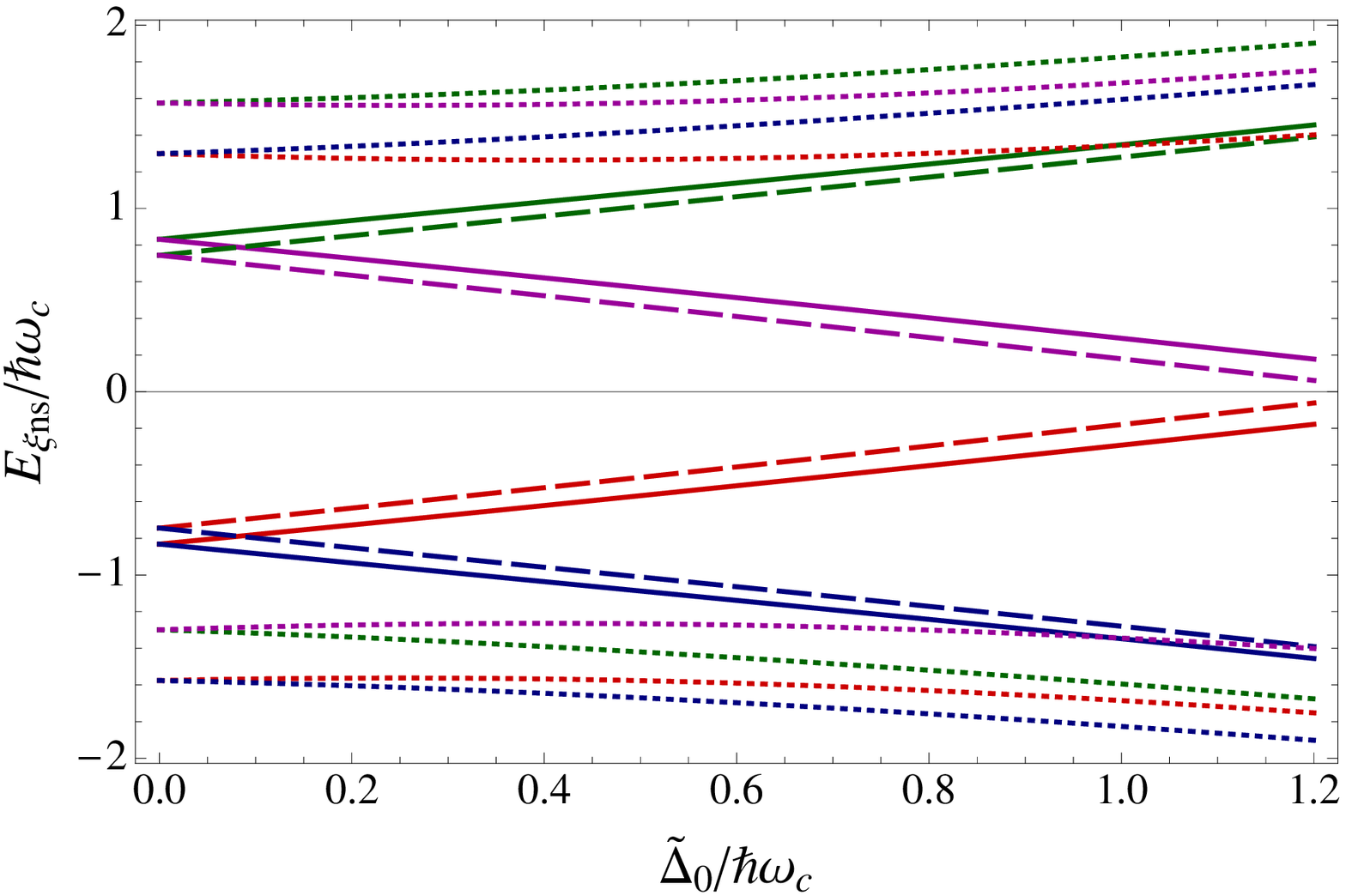}
\includegraphics[width=.32\textwidth]{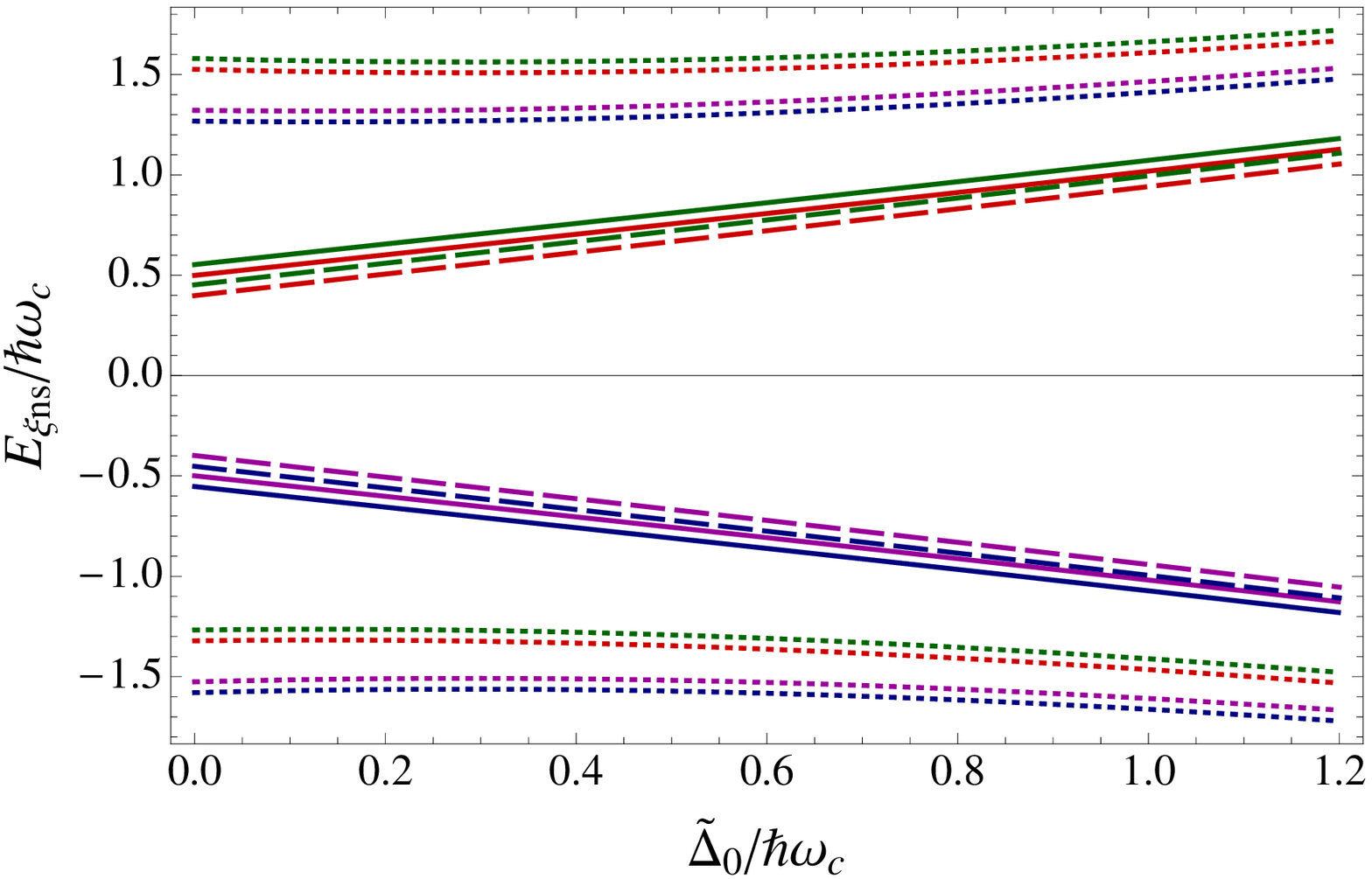}
\includegraphics[width=.32\textwidth]{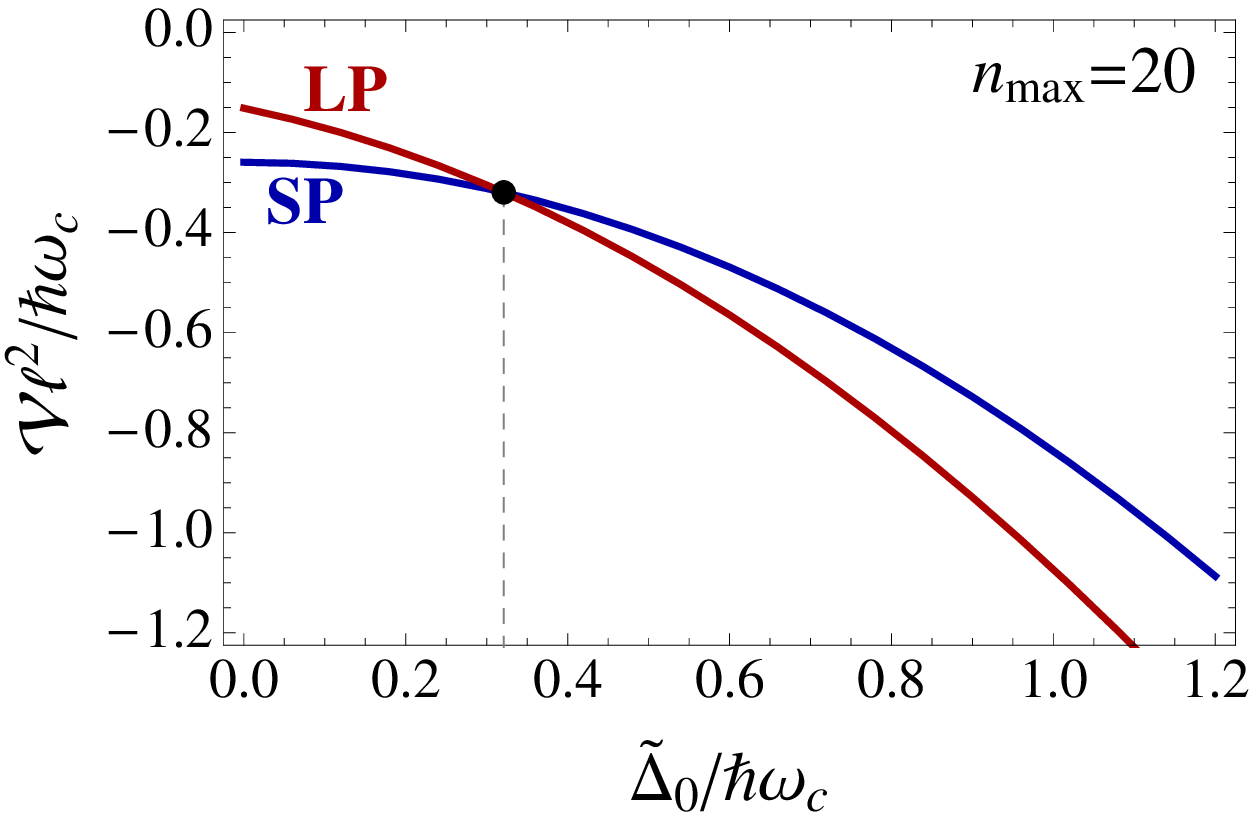}
\caption{(Color online) Quasiparticle energies in the lowest three Landau levels, $n=0$ (solid lines), $n=1$
(dashed lines) and $n=2$ (dotted lines), for the SP (left panel) and LP (middle panel) solutions
obtained in the approximation with static screening at $B= 2\,\mbox{T}$.  Colors of
the lines correspond to specific values of quantum numbers $(\xi,s)$: red to $(-,\downarrow)$,
green to $(-,\uparrow)$, blue to $(+,\downarrow)$, and purple to $(+,\uparrow)$. The right
panel shows the free energies of the same solutions as functions of $\tilde{\Delta}_0$.}
\label{Type_I_II-stat-screen}
\end{center}
\end{figure*}

Let us start our analysis by considering a simple benchmark case with $B= 2\,\mbox{T}$ ($n_{\rm max}= 20$)
and $\kappa=2$. For a fixed value of the voltage imbalance $\tilde{\Delta}_0$,
we typically find two types of competing solutions: spin-polarized (SP) and
layer-polarized (LP) ones. The examples of both types of solutions are shown in
Fig.~\ref{pars-vs-n-static}, which are obtained at a fixed value of $\tilde{\Delta}_{0}\approx
0.3 \hbar\omega_c$, which is close to the critical value $\tilde{\Delta}_{0,{\rm cr}} \approx 0.32 \hbar\omega_c$  
separating the SP and LP phases (see Fig.~\ref{Type_I_II-stat-screen} and its discussion below).   
Note the different energy scales on the left and right panels in Fig.~\ref{pars-vs-n-static}.

As one can see from this figure, the QHF parameters ($\mu_{ns}$, $\tilde{\mu}_{ns}$) and the
MC ones ($\Delta_{ns}$, $\tilde{\Delta}_{n,s}$) {\it coexist} in all Landau levels.
The Haldane mass $\Delta_{ns}$ and the 
chemical potential $\mu_{ns}$ have different signs for up and down spins in the SP solution. The values of 
$|\Delta_{0,s}| \simeq 0.83\hbar\omega_c$ and 
$|\Delta_{1,s}| \simeq 0.74\hbar\omega_c$ in the LLL are essentially larger than those for higher LLs with $n \geq 2$, 
which are slowly decreasing with increasing $n$: from $|\Delta_{2,s}| \simeq 0.25\hbar\omega_c$ down to 
$|\Delta_{20,s}| \simeq 0.13\hbar\omega_c$. Its QHF counterpart $\mu_{ns}$ behaves similarly: 
$|\mu_{2,s}| \simeq 0.14\hbar\omega_c$ and $|\mu_{20,s}| \simeq 0.11\hbar\omega_c$. 
(Recall that $\hbar\omega_c \simeq 4.30 \, \mbox{meV}$ for $B = 2\,\mbox{T}$.)

As to the dynamical voltage imbalance (Dirac mass) $\tilde{\Delta}_{ns}$, it is suppressed with 
respect to the bare voltage 
$\tilde{\Delta}_{0}\approx 0.3 \hbar\omega_c$ used in this figure: 
$\tilde{\Delta}_{0,s} \simeq 0.16 \hbar\omega_c$ and 
$\tilde{\Delta}_{1,s} \simeq 0.16 \hbar\omega_c$
in the LLL, while 
$\tilde{\Delta}_{2,s} \simeq 0.23 \hbar\omega_c$ and 
$\tilde{\Delta}_{20,s} \simeq 0.15 \hbar\omega_c$. 
The value of its QHF counterpart $\tilde{\mu}_{ns}$ is very small. It starts from
$\tilde{\mu}_{2,s} \simeq 0.009 \hbar\omega_c$ and decreases down to the values of order 
$10^{-4}\hbar\omega_c$ at large $n$.
Thus, taking into account the dispersion relations in Eqs.~(\ref{LLLn}) and (\ref{2n}), we conclude 
that, as expected, the splitting of the levels with opposite spins is responsible for generating a gap in the SP solution (see
Fig.~\ref{Type_I_II-stat-screen} and its discussion below).

In the LP solution, the values $|\Delta_{ns}|$ and  $|\mu_{ns}|$ are small. In fact,
while $|\Delta_{0s}|$ and $|\Delta_{1s}|$ are equal to the Zeeman energy $Z = 0.027 \hbar\omega_c$,
all $\Delta_{ns}$ with $n \geq 2$ vanish. The chemical potential $|\mu_{ns}|$ is equal to the Zeeman energy
for all $n$. As to the parameters 
$\tilde{\Delta}_{ns}$ and $\tilde{\mu}_{ns}$, the values of the former in the LLL are significantly
larger than those in the higher LLs: 
$\tilde{\Delta}_{0, s} = 0.68\hbar\omega_c$, 
$\tilde{\Delta}_{1,s} = 0.59\hbar\omega_c$, while 
$\tilde{\Delta}_{2,s} = 0.07\hbar\omega_c$ and 
$\tilde{\Delta}_{20,s} = 0.007\hbar\omega_c$. 
Its QHF counterpart parameters $\tilde{\mu}_{n,s}$ are 
$\tilde{\mu}_{2,s} = - 0.12\hbar\omega_c$, 
$\tilde{\mu}_{3,s} = - 0.10\hbar\omega_c$, and 
$\tilde{\mu}_{20,s} = - 0.08\hbar\omega_c$, whose absolute values are considerably larger than 
the Zeeman energy. From the dispersion relations in Eqs.~(\ref{LLLn}) and (\ref{2n}), we conclude 
that the splitting of the levels assigned to different valleys, $K$ ($\xi =1$) 
and $K^{\prime}$ ($\xi = -1$), is responsible for generating a gap in the LP solution (recall that 
the valley and layer degrees of freedom are not independent in the LLL).

Although the values of the QHF and MC parameters are different at other values of the 
bare voltage imbalance $\tilde{\Delta}_0$, the main characteristics of their dependence on the Landau 
index $n$ remain qualitatively similar. Instead of showing the QHF and MC parameters as functions of 
Landau level index $n$ for other values of $\tilde{\Delta}_0$, it is convenient 
to summarize the $\tilde{\Delta}_0$ dependence of the SP- and LP-type solutions by presenting the 
spectra of the first few low-energy states in Fig.~\ref{Type_I_II-stat-screen}. The first two panels show 
the energies of the first few Landau levels for the SP and LP solutions. The free energies are compared 
in the right panel of the same figure.

As we see, the spectrum of the LLL with $n = 0$ and $n = 1$ is qualitatively different 
from that of the $n = 2$ LL. The roots of this difference are in the form of the spectrum in
the bilayer model without interactions.\cite{Falko} While the energies of LLL states $E_{\xi 0 s}$ 
and $E_{\xi 1 s}$ equal zero, there are positive and negative energy bands $E_{\xi n s}^{\pm} = 
\pm\sqrt{n(n-1)}\hbar\omega_c$ for each of the higher LLs with $n \geq 2$.

The SP solution is the ground state at small values of $\tilde{\Delta}_0$, while the LP one becomes 
the ground state at large values of $\tilde{\Delta}_0$. The first order phase transition\cite{GGM2,GGJM} 
from one regime to the other occurs at the critical value $\tilde{\Delta}_{0,{\rm cr}} \approx 0.32 \hbar\omega_c$. 
In terms of the applied electric field, this is equivalent to $E_{\perp,{\rm cr}}\approx 7.87\,\mbox{mV/nm}$ 
(at fixed value $B= 2\,\mbox{T}$), which is somewhat smaller than typical values measured in
the experiment.\cite{Weitz,Kim,Freitag,Velasco}

It is instructive to compare the gaps in the energy spectra for both the SP and LP solutions at a fixed 
value of magnetic field, $B = 2\,\mbox{T}$. At $\tilde{\Delta}_0 = 0$ the energy gap for the SP solution is 
$E_{\rm gap}^{\rm SP} = E^{\rm SP}_{+,1,\uparrow} - E^{\rm SP}_{-,1,\downarrow} \simeq 1.49 \hbar\omega_c$, 
which is considerably larger than the corresponding gap for the LP solution, 
$E_{\rm gap}^{\rm LP} =  E^{\rm LP}_{-,1,\downarrow} - E^{\rm LP}_{+,1,\uparrow} \simeq 0.80 \hbar\omega_c$. 
At the critical value $\tilde{\Delta}_{0,\rm cr} \simeq 0.32\hbar\omega_c$, however, both gaps happen to be 
approximately equal to $1.14\hbar\omega_c$. These values are also comparable to those in some 
experiments,\cite{Weitz,Martin} but are considerably smaller than the recent experimental values in 
suspended bilayer graphene with a high mobility.\cite{Velasco}

From a theoretical viewpoint, it is interesting to investigate the dependence of the 
main results on $n_{\rm max}$, which plays the role of the high-energy cutoff parameter. While 
the qualitative competition between the SP and LP states remains the same, the critical values 
$E_{\perp,{\rm cr}}$ decrease slightly with decreasing the number of Landau levels (i.e., shrinking 
the available phase space). The actual critical values of the applied electric field (at $B= 2\,\mbox{T}$) 
for several different values of the Landau level cutoff parameter $n_{\rm max}$ are given in 
Table~\ref{tab-E-crit}. These data show that the critical value of $E_{\perp,{\rm cr}}$ may 
have a sizable uncertainty, associated with the choice of $n_{\rm max}$. Indeed, its approximate 
physical value is determined only semi-rigorously as the number of Landau levels below the energy 
cutoff $\Lambda\simeq \gamma_1/4$.

\begin{table*}[ht]
\caption{Critical values of $\tilde{\Delta}_{0}$ and $E_{\perp}$ for several choices of the cutoff parameter
$n_{\rm max}$ in the approximation with static screening at $B= 2\,\mbox{T}$ and $\kappa=2$.}
\begin{ruledtabular}
\begin{tabular}{l|llllllll}
$n_{\rm max}$ & $1$ & $2$ & $3$ & $4$ & $5$ & $6$ & $10$ & $20$ \\
\hline
$\tilde{\Delta}_{0,{\rm cr}}/\hbar\omega_{c} $      & 0.204 & 0.234 & 0.250 & 0.260 & 0.268 & 0.275 & 0.293 & 0.318 \\
$E_{\perp,{\rm cr}}~\mbox{[mV/nm]}$                 & 5.02 & 5.75 & 6.14 & 6.40 & 6.60 & 6.75 & 7.20 & 7.83
\label{tab-E-crit}
\end{tabular}
\end{ruledtabular}
\end{table*}

Now, let us discuss the dependence of our numerical results on the magnetic field. Neglecting
a weak dependence of the interaction coefficients $I_{n,n^\prime}$ on the magnetic field $B$ in the
static approximation, one may claim that the full dependence of the dynamical parameters on $B$
can be completely restored from simple scaling arguments. As seen from the gap equations
(\ref{dyn-gap-eq-n=0}) -- (\ref{dyn-gap-eq-n>1B}), the main dependence on the magnetic
field comes through the overall factor $\hbar\omega_c$ on the right hand side of all equations.
(Recall that, in the static approximation, the coefficient functions $I_{n^\prime,n}$ are
energy independent.) This dependence on $\hbar\omega_c$ can be removed by
introducing dimensionless energy parameters, measured in units of $\hbar\omega_c$.
If this were the only dependence on the magnetic field, the results for all dimensionless
ratios, such as $\mu_{ns}/\hbar\omega_c$, $\tilde{\mu}_{ns}/\hbar\omega_c$, $\Delta_{ns}/\hbar\omega_c$,
$\tilde{\Delta}_{ns}/\hbar\omega_c$ and all energies $E_{\xi n s}/\hbar\omega_c$ would be the same
for any value of $B$.

An additional weak dependence on the magnetic field enters the gap equations indirectly
through the cutoff parameter $n_{\rm max}$ and through the $x$-parameter, defined after
Eq.~(\ref{V-eff-static}). The latter appears in the definition of the coefficient functions
$I_{n^\prime,n}$ even in the static approximation. However, the corresponding
dependence on the magnetic field is relatively weak. When the value of $B$ increases by two
orders of magnitude (from $0.1\,\mbox{T}$ to $10\,\mbox{T}$), the numerical values of all
off-diagonal coefficients $I_{n^\prime,n}$ ($n^\prime\neq n$) change only by a few
percent. There is a stronger dependence on the field in the diagonal coefficient $I_{n,n}$,
but even this one changes only by about a factor of 2.

In order to see the deviations from the linear scaling laws, we did perform a careful numerical
dependence of the results on the magnetic field. Note that, in such a study, one has to adjust
properly the maximum value of the Landau levels kept in the gap equations.
Recalling that $n_{\rm max}= k_B\Lambda/(\hbar \omega_c) \simeq 40/(B\,\mbox{[T]})$,
we will consider several values of magnetic fields (e.g., $2 \,\mbox{T}$, $4 \,\mbox{T}$,
$5 \,\mbox{T}$, $8 \,\mbox{T}$) which lead to integer values of $n_{\rm max}$. 
The best fits to the critical lines are given by the following expressions:
\begin{eqnarray}
\kappa=1: &&  E_{\perp,{\rm cr}}\simeq  \left(3.82 + 5.98 B~\mbox{[T]}\right)~\mbox{mV/nm},\\
\label{linear-static-offset}
\kappa=2: &&  E_{\perp,{\rm cr}}\simeq  \left(1.69 + 3.07 B~\mbox{[T]}\right)~\mbox{mV/nm}.
\end{eqnarray}
One of the qualitative deviations from the linear scaling is the appearance of a small positive
offset in the dependence of the critical value $E_{\perp,{\rm cr}}$ on the magnetic field.
It is noticeable that such an offset would be absent without Landau level mixing.
Although strictly speaking we cannot consider the limit of vanishingly small magnetic fields using
the present approach, the interpolation of this result qualitatively agrees with the experimental
data\cite{Weitz,Kim,Freitag,Velasco} showing that the SP state extends all the way to $B=0$ at sufficiently small electric fields.
This seems to suggest that it can be the ground state of bilayer graphene at the neutral point in the absence
of external fields.

Comparing these results with the earlier studies without Landau level mixing, we see
that the size of the energy gaps in the SP and LP type ground state as well as the critical values
of the applied electric field substantially increased. However, even the inclusion of Landau mixing 
did not resolve completely the discrepancy with the corresponding experimental values in clean 
samples,\cite{Velasco} which appear to be still larger than our predictions in a model with static 
screening. It is critical, therefore, to go beyond the static approximation.

\subsection{Dynamical screening}
\label{DynamicalApproximation}

Now let us turn to the analysis of the gap equations with dynamical screening. Computationally,
this is much harder than the static case. It is a nontrivial energy dependence of the coefficient
functions $I_{n^{\prime},n}(E)$ that makes the numerical computations considerably
slower. Also, a highly nonlinear nature of the gap equations makes finding numerical
solutions much harder. Before even solving the gap equations, in fact, one first needs to compile
all coefficient functions $I_{n^{\prime},n}(E)$ with $n^{\prime},n\leq n_{\rm max}$ using 
their definition in Eq.~(\ref{Innprime}). Fortunately, the task is somewhat simplified by the observation that
each of these functions can be fitted quite well with a simple Pad\'{e} approximant of order [2/1],
see Eq.~(\ref{Innprime-Pade}). The numerical values of the coefficients for the fits of
$I_{n^{\prime},n}(E)$ with $n^{\prime},n\leq 10$ are listed in Table~\ref{tab-a-b-c-d}
in Appendix~\ref{Tables}.

\begin{figure*}
\begin{center}
\includegraphics[width=.8\textwidth]{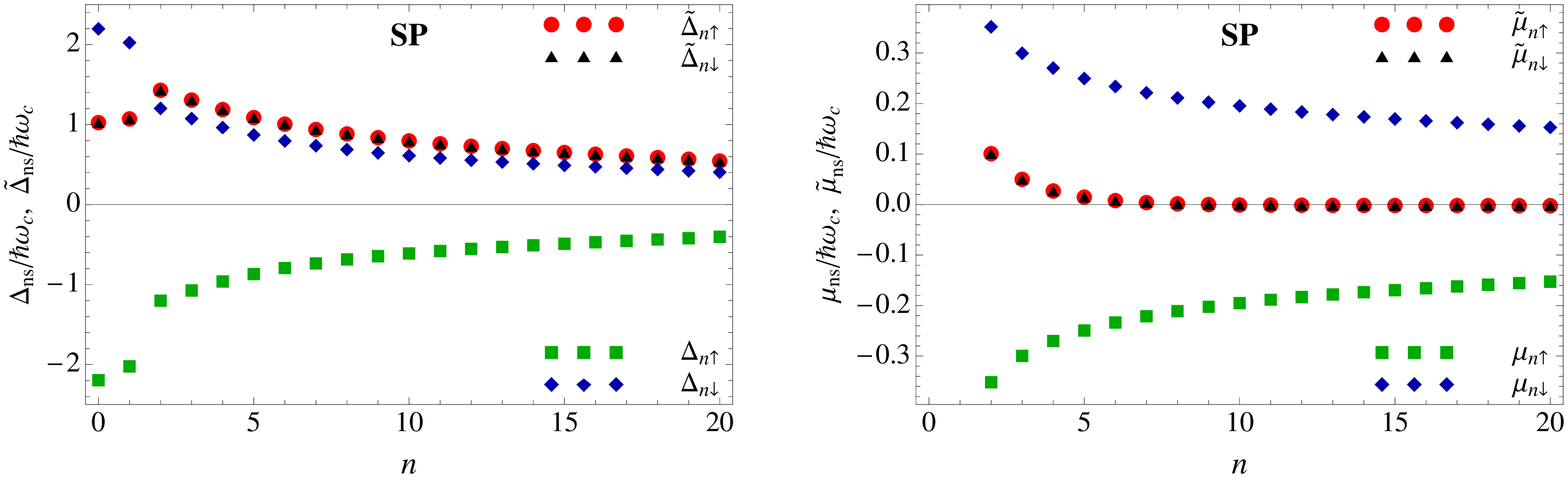}\\
\includegraphics[width=.8\textwidth]{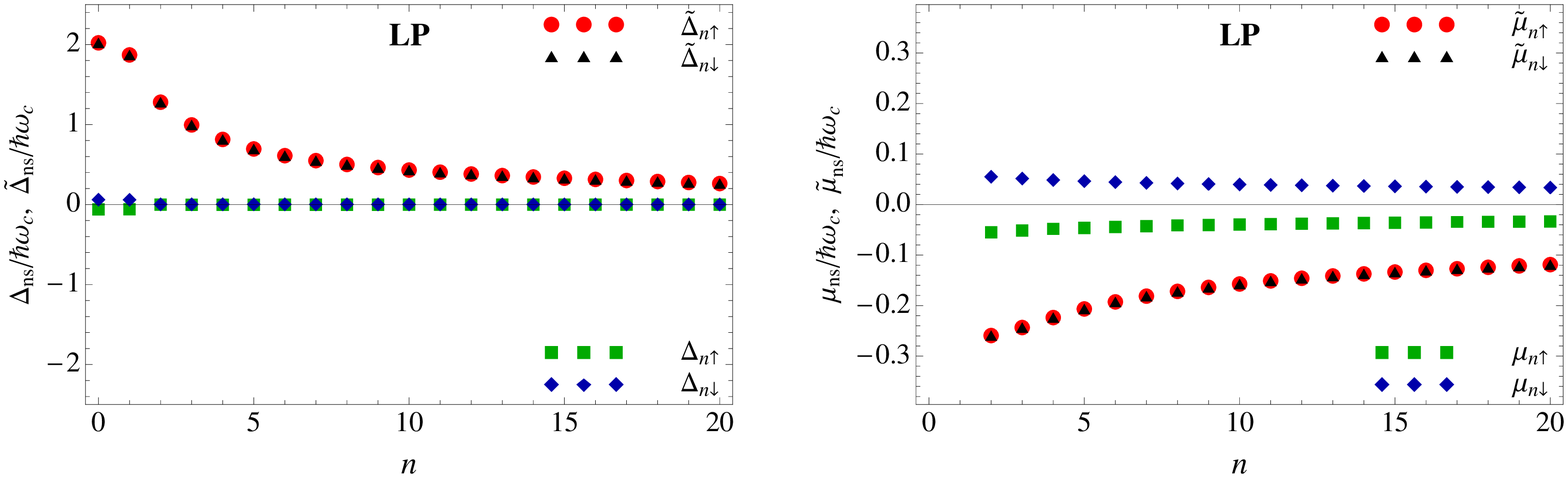}
\caption{(Color online) The dependence of the dynamical parameters on the Landau level index
$n$ for the SP (top panels) and LP (bottom panels) solutions at a fixed value of 
$\tilde{\Delta}_{0}\approx \hbar\omega_c$ and $\kappa=2$. The numerical results are obtained
in the approximation with $n_{\rm max}=20$ and dynamic screening. The magnetic field is $B= 2\,\mbox{T}$.}
\label{pars-vs-n}
\end{center}
\end{figure*}

Let us start from considering some key results in the same benchmark case of a fixed
magnetic field, $B= 2\,\mbox{T}$. 
By reviewing the functional dependence of the coefficients $I_{n^{\prime},n}(E)$, compiled
numerically, we find that these functions grow substantially with energy. For example, typical
values of the coefficients $I_{n^{\prime},n}(E)$  with $n^{\prime},n\leq 10$ at $E=\hbar\omega_c$
appear to be about $20\%$ to $90\%$ larger then their values at $E=0$ (i.e., static
screening approximation). From this fact alone, it is natural to expect that the dynamical
parameters as well as the associated energy gaps in the spectra should be larger in the
approximation with dynamic screening. This is indeed what we see. 
The examples of the SP and LP solutions are shown in Fig.~\ref{pars-vs-n}, which are obtained at 
a fixed value of $\tilde{\Delta}_{0}\approx 1.02\hbar\omega_c\simeq 4.39\, \mbox{meV}$ that is close to 
the critical value $\tilde{\Delta}_{0,{\rm cr}}\simeq 0.99\hbar\omega_c$
(note the different energy scales on the left and right panels in Fig.~\ref{pars-vs-n}). This
figure illustrates that, as in the case of the static screening, the MC and QHF
dynamical parameters necessarily coexist.

Although the MC and QHF parameters in the case of the dynamical screening are considerably larger than
in the case of the static screening, their functional dependence on the Landau index $n$ and some other important 
characteristics are similar.  
As we see from Fig.~\ref{pars-vs-n}, the Haldane mass $\Delta_{ns}$ and the chemical potential $\mu_{ns}$ have
different signs for up and down spins in the SP state. The values of $|\Delta_{0,s}| \simeq 2.20\hbar\omega_c$ and 
$|\Delta_{1,s}| \simeq 2.02\hbar\omega_c$ in the LLL are substantially larger than those in higher LLs with $n \geq 2$, 
which are slowly decreasing with $n$ from $|\Delta_{2,s}| \simeq 1.20\hbar\omega_c$ down to 
$|\Delta_{20,s}| \simeq 0.40\hbar\omega_c$. Its QHF counterpart $\mu_{ns}$ behaves similarly, decreasing 
from $|\mu_{2,s}| \simeq 0.35\hbar\omega_c$ to $|\mu_{20,s}| \simeq 0.15\hbar\omega_c$.

\begin{figure*}
\begin{center}
\includegraphics[width=.32\textwidth]{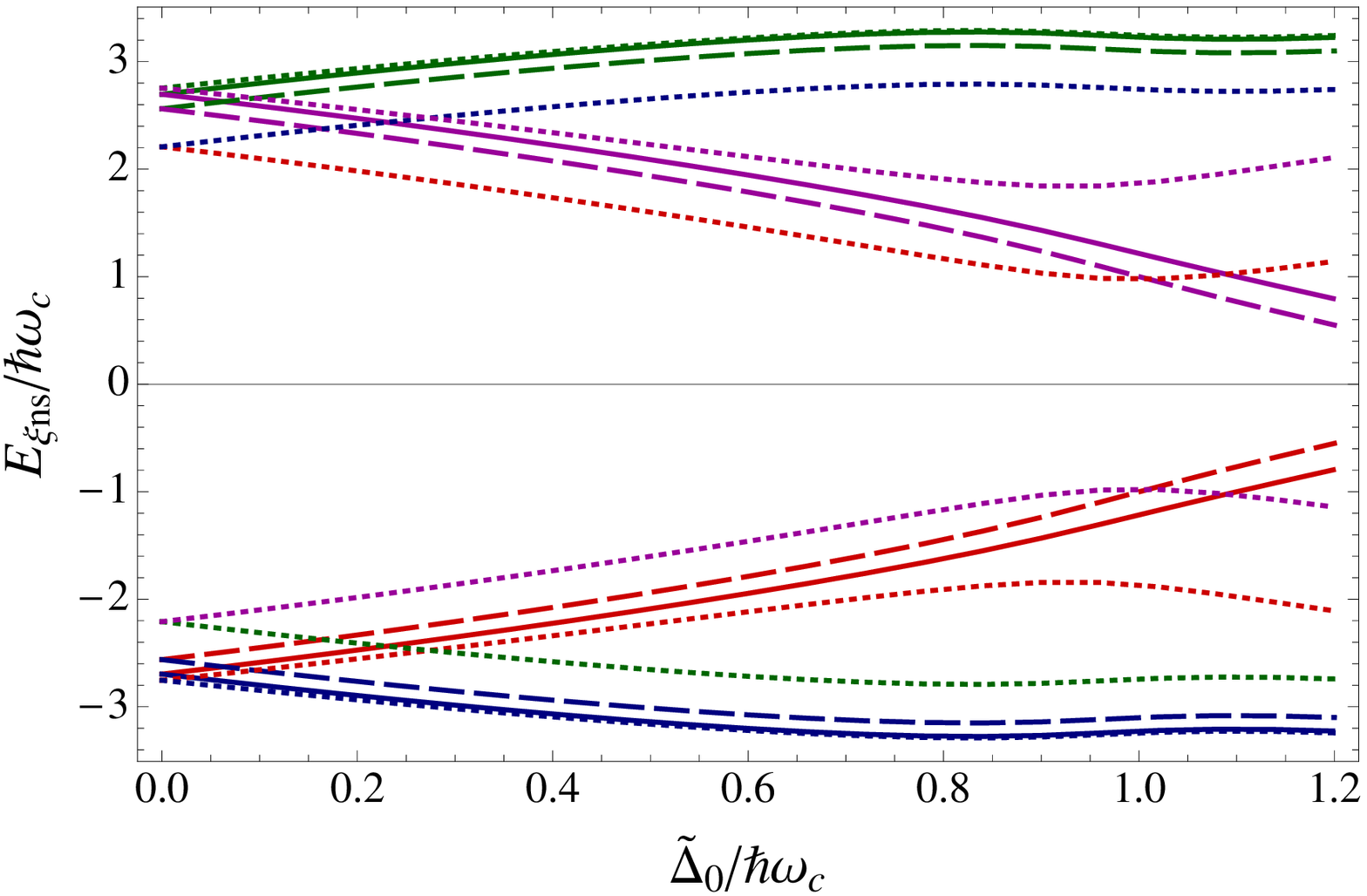}
\includegraphics[width=.32\textwidth]{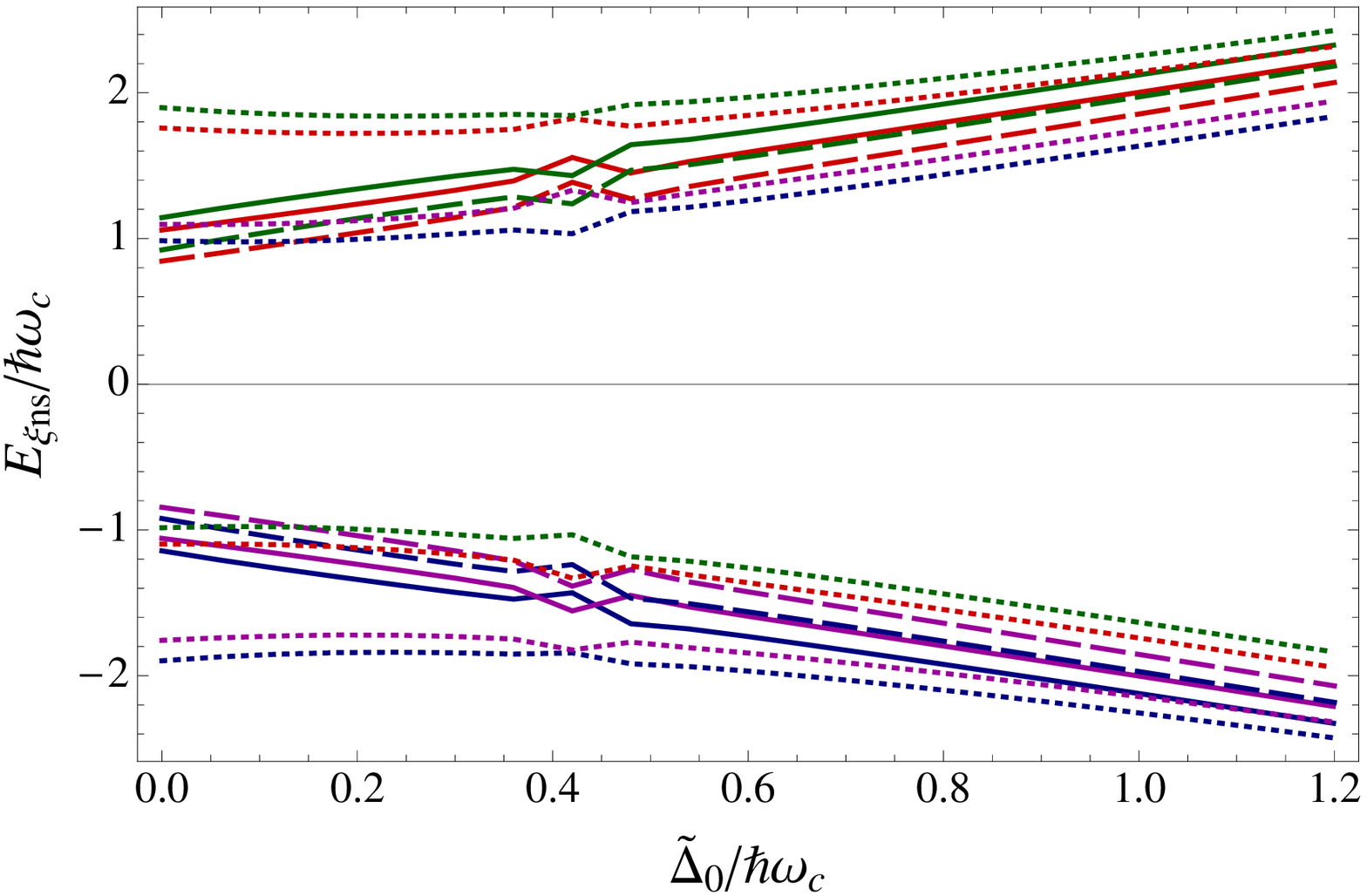}
\includegraphics[width=.32\textwidth]{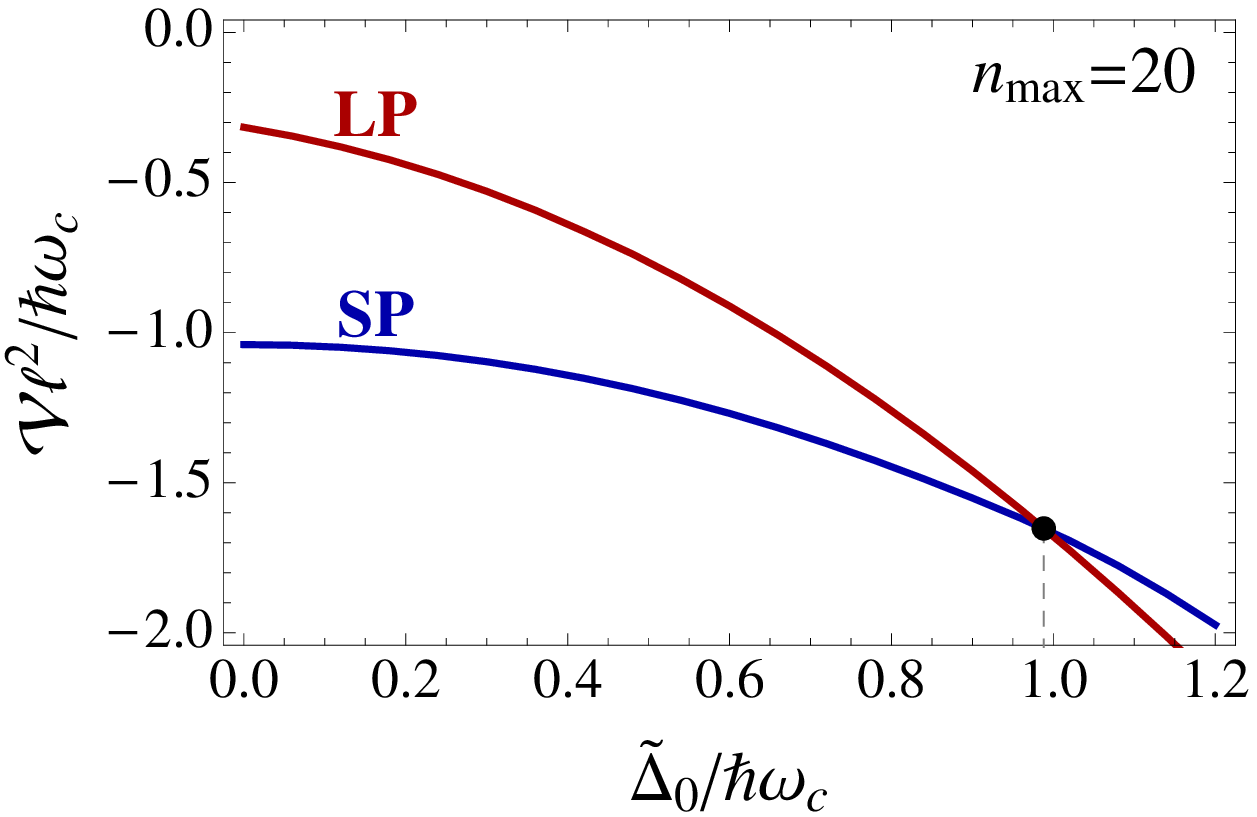}
\caption{(Color online) Quasiparticle energies in the lowest three Landau levels, $n=0$ (solid lines), $n=1$
(dashed lines) and $n=2$ (dotted lines), for the SP (left panel) and LP (middle panel) solutions
obtained in the approximation with {\em dynamic} screening at $B= 2\,\mbox{T}$. Colors of the 
lines correspond to specific values of quantum numbers $(\xi,s)$: red to $(-,\downarrow)$, green to $(-,\uparrow)$,
blue to $(+,\downarrow)$, and purple to $(+,\uparrow)$.
The right panel shows the free energies of the same solutions as functions of $\tilde{\Delta}_0$.}
\label{Type_I_II-dynamic-screen}
\end{center}
\end{figure*}

As to the dynamical voltage imbalance (Dirac mass) $\tilde{\Delta}_{ns}$, unlike the static case,  
its values in the LLL and $n=2$ LL are not suppressed with respect to the bare voltage 
$\tilde{\Delta}_{0}\approx 1.02 \hbar\omega_c$. Namely, $\tilde{\Delta}_{0,s} \simeq 1.02 \hbar\omega_c$ and 
$\tilde{\Delta}_{1,s} \simeq 1.07 \hbar\omega_c$ in the LLL, while $\tilde{\Delta}_{2,s} \simeq 1.43 \hbar\omega_c$. Its 
value at the largest $n= 20$ is $\tilde{\Delta}_{20,s} \simeq 0.54 \hbar\omega_c$. 
As in the static case, the values of its QHF counterpart $\tilde{\mu}_{ns}$ are much smaller, starting from
$\tilde{\mu}_{2,s} \simeq 0.10 \hbar\omega_c$ and decreasing down to the values of order $10^{-3}\hbar\omega_c$ at large $n$.
Therefore, we conclude that the splitting of 
the levels with opposite spins is responsible for generating a gap in the SP solution (see
also Fig.~\ref{Type_I_II-dynamic-screen} and its discussion below).

In the LP solution, the values $|\Delta_{ns}|$ and  $|\mu_{ns}|$ are small, although
unlike the static screening case, their values in the LLL are larger that the Zeeman energy
$Z = 0.027\hbar\omega_c$. In fact, $|\Delta_{0,s}|\simeq 0.06\hbar\omega_c$ and $|\Delta_{1,s}|\simeq 0.06\hbar\omega_c$, while 
all $\Delta_{ns}$ with $n \geq 2$ are of the order $10^{-3}\hbar\omega_c$. The chemical 
potential $|\mu_{ns}|$ slowly decreases from $|\mu_{2,s}| \simeq 0.055\hbar\omega_c$ to 
$|\mu_{20,s}| \simeq 0.001\hbar\omega_c$. As to the parameters 
$\tilde{\Delta}_{ns}$ and $\tilde{\mu}_{ns}$, the values of the former in the LLL are significantly
larger than those in the higher LLs: 
$\tilde{\Delta}_{0, s} \simeq 2.08\hbar\omega_c$, 
$\tilde{\Delta}_{1,s} \simeq 1.93\hbar\omega_c$, while 
$\tilde{\Delta}_{2,s} \simeq 1.36\hbar\omega_c$ and 
$\tilde{\Delta}_{20,s} \simeq 0.30\hbar\omega_c$. Its QHF counterpart $\tilde{\mu}_{n,s}$ is
$\tilde{\mu}_{2,s} \simeq - 0.25\hbar\omega_c$, 
$\tilde{\mu}_{3,s} \simeq - 0.24\hbar\omega_c$, and 
$\tilde{\mu}_{20,s} \simeq - 0.12\hbar\omega_c$, whose absolute values are considerably larger than 
the Zeeman energy. From the dispersion relations in Eqs.~(\ref{LLLn}) and (\ref{2n}), we conclude 
that the splitting of the levels assigned to different valleys, $\xi = \pm1$, is responsible for generating 
a gap in the LP solution.

The energy spectra of the low-energy states are presented in Fig.~\ref{Type_I_II-dynamic-screen}.
A few comments are in order here. In addition to the expected large values of the gaps, we also 
find that (i) the energies have a rather complicated dependence on the
voltage imbalance $\tilde{\Delta}_0$, which substantially deviates from a linear dependence,
(ii) there are several points of level crossings and (iii) a nonstandard order of the Landau levels with
the lowest energy state being the $n=2$ Landau level. The reason of the last phenomenon
is connected with the following feature: As one can see in Fig.~\ref{pars-vs-n}, the values of
some dynamical parameters in the LLL are much larger than those in the $n=2$ Landau level.
Interestingly, this feature takes place for both the SP and LP solutions.

\begin{figure}
\begin{center}
\includegraphics[width=.4\textwidth]{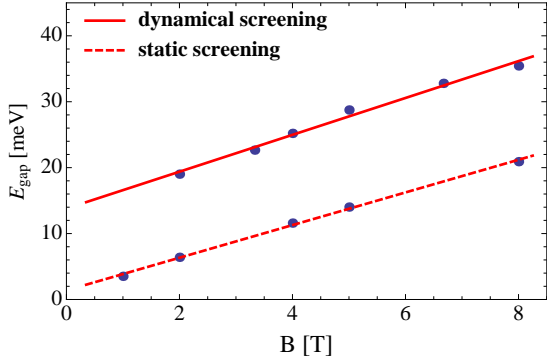}
\caption{(Color online) The energy gap $E_{\rm gap}$ in the SP state as a function of the 
magnetic field at fixed $\tilde{\Delta}_0= 0$.}
\label{gap-vs-B}
\end{center}
\end{figure}

The energy gap in the SP ground state is given by 
$E^{\rm SP}_{\rm gap} = E_{-,2,\downarrow} - E_{+,2,\uparrow}$. At $\tilde{\Delta}_0= 0$, the 
corresponding value of the gap is $E^{\rm SP}_{\rm gap} \simeq 4.42\hbar\omega_c$ 
at $B=2\,\mbox{T}$. This gap grows with increasing magnetic field, and the corresponding 
dependence is approximately a linear function of $B$. Our numerical results and linear fits 
are shown in Fig.~\ref{gap-vs-B} for the cases of dynamical as well as static screening.

\begin{figure}
\begin{center}
\includegraphics[width=.4\textwidth]{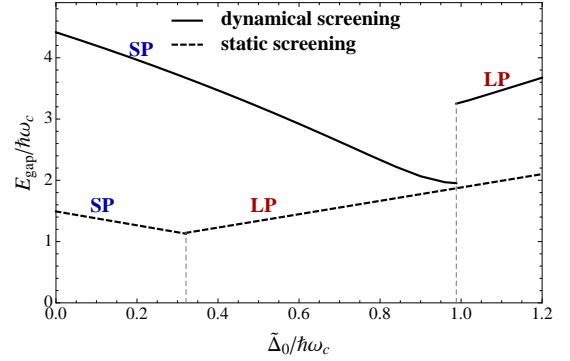}
\caption{(Color online) The energy gap $E_{\rm gap}$ as a function of $\tilde{\Delta}_0$ at fixed 
$B= 2\,\mbox{T}$ in the case of dynamical and static screening.}
\label{fig-E-gap}
\end{center}
\end{figure}

At the critical point, $\Delta_{0,{\rm cr}} \simeq 0.99\hbar\omega_c$, the SP gap 
$E^{\rm SP}_{\rm gap}$ and the LP one, $E^{\rm LP}_{\rm gap} = E_{+,2,\uparrow}-E_{-,2,\uparrow}$, 
take different values: $E^{\rm SP}_{\rm gap} \simeq 1.95\hbar\omega_c$ and $E^{\rm LP}_{\rm gap} \simeq 3.24\hbar\omega_c$. 
In other words there is a jump in the gap at the critical point shown in Fig.~\ref{fig-E-gap}.
For comparison, the gap in the case of the static screening is also shown in this figure. 
As one can see, at the critical point it has a kink that is smoother than a jump singularity. 
As discussed in Sec.~\ref{Discussion} below, the presence of such singularities at the critical 
point is relevant for understanding the behavior of the conductivity observed in experiment.\cite{Weitz}

Another consequence of the approximation with the dynamical screening is a substantial enhancement of
the critical value of the applied electric field, at which the phase transition from SP to LP states occurs.
For the results presented in Fig.~\ref{Type_I_II-dynamic-screen}, for example, the critical value is
$E_{\perp,{\rm cr}}\approx 24.31\,\mbox{mV/nm}$ (as seen from the figure for the free energy,
the corresponding voltage imbalance is $\tilde{\Delta}_{0,{\rm cr}} \approx 0.99 \hbar\omega_c$). In order to
appreciate the effect of Landau level mixing, we present the numerical values of $\tilde{\Delta}_{0}$
and $E_{\perp}$ for several choices of the cutoff parameter $n_{\rm max}$ in Table~\ref{tab-critical}.

\begin{table*}[ht]
\caption{Critical values of $\tilde{\Delta}_{0}$ and $E_{\perp}$ for several choices of the cutoff parameter
$n_{\rm max}$ in the approximation with dynamical screening at $B= 2\,\mbox{T}$ and $\kappa=2$.}
\begin{ruledtabular}
\begin{tabular}{l|llllllll}
$n_{\rm max}$ & $1$ & $2$ & $3$ & $4$ & $5$ & $6$ & $10$ & $20$ \\
\hline
 $\tilde{\Delta}_{0,{\rm cr}}/\hbar\omega_{c} $      & 0.241 & 0.336 & 0.412 & 0.476 & 0.530 & 0.578 & 0.732 & 0.988 \\
 $E_{\perp,{\rm cr}}~\mbox{[mV/nm]}$ & 5.92 & 8.27 & 10.2 & 11.7 & 13.0 & 14.2 & 18.0 & 24.3
\label{tab-critical}
\end{tabular}
\end{ruledtabular}
\end{table*}

\begin{figure}
\begin{center}
\includegraphics[width=.4\textwidth]{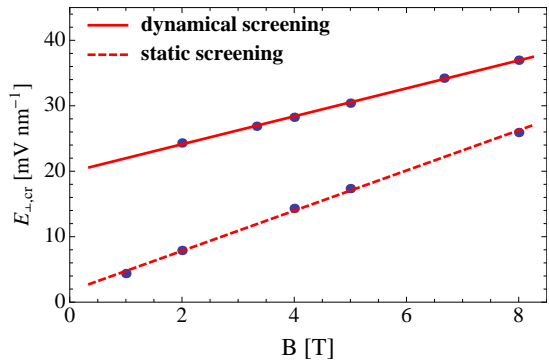}
\caption{(Color online) The dependence of the critical value of $E_{\perp}$ on the 
magnetic field in the case of $\kappa=2$.}
\label{E-critical-vs-B}
\end{center}
\end{figure}

The corresponding numerical results for magnetic field dependence of the critical value of $E_{\perp}$ 
are shown in Fig.~\ref{E-critical-vs-B}. Just like in the static case, we adjust the
maximum value of the Landau levels as follows: $n_{\rm max}= k_{B}\Lambda/\hbar\omega_c(B)$, where $\Lambda
=1000\,\mbox{K}$ is a fixed cutoff. The data can be well approximated by the following linear 
dependence with a nonzero offset:
\begin{equation}
E_{\perp,{\rm cr}}\simeq  \left(19.85 + 2.13 B~\mbox{[T]}\right)~\mbox{mV/nm},
\quad (\Lambda=1000\,\mbox{K}).
\label{linear-offset}
\end{equation}
Comparing this expression with that in Eq.~(\ref{linear-static-offset}), we conclude that
the dynamical $E_{\perp,{\rm cr}}$ is considerably larger than the static one.

\section{Discussion}
\label{Discussion}

The two ingredients, the Landau level mixing and the dynamical screening, play an important 
role in the dynamics in bilayer graphene in a magnetic field.\cite{collective} In fact, their role is more important 
than that in monolayer graphene (compare with Ref.~\onlinecite{GGMS11.05}). This is a reflection 
of the general fact that the Coulomb interaction in bilayer graphene plays a more profound role. The latter is 
the consequence of a much smaller characteristic energy scale 
$\hbar \omega_{c} \simeq 2.15 B\, [\mbox{T}]\, \mbox{meV}$ in bilayer 
graphene as compared to the Landau energy scale 
$\varepsilon_{\ell}\simeq 26 \sqrt{B\, [\mbox{T}]}\, \mbox{meV}$ in monolayer.

In this study, we derived and solved a complete set of gap equations with Landau mixing
in the Hartree-Fock approximation with the frequency-dependent polarization function
calculated at the charge neutrality point. The competition between the
spin-polarized and layer polarized states is studied in detail by varying the applied
electric field (or, equivalently, the voltage imbalance between the top and bottom layers)
and the strength of the magnetic field. 

It was found that the critical value of the applied electric field is a linear function of the
magnetic field, i.e., $E_{\perp,{\rm cr}}=E_{\perp}^{\rm off}+a B$, where $E_{\perp}^{\rm off}$
is an offset electric field and $a$ is the slope. The offset electric field and energy gaps
substantially increase with the inclusion of dynamical screening compared to the case of
static screening. The solutions to the gap equations clearly demonstrate that the QHF and
MS order parameters necessarily coexist.

We demonstrated that the use of the static screening approximation badly
underestimates the strength of the direct Coulomb interaction. Taking into account
the effects of dynamic screening leads to energy gaps that are about a factor of 2
to 3 larger than those in the approximation with static screening. The increased
values of the gaps as well as more pronounced nonlinearity of the gap equations
result in a rather non-trivial energy spectrum. In particular, we find that the energy
of the $n=2$ Landau level becomes smaller than the $n=0,1$ (``lowest'') Landau
levels.

Let us now compare the theoretical model results with the experimental data in 
Refs.~\onlinecite{Weitz,Martin,Velasco}. One of the most noticeable features of the 
present model is the approximate linear dependences of the energy gap $E_{\rm gap}$ and 
the critical electric field $E_{\perp,{\rm cr}}$ on the magnetic field. We find not only that these 
functions are linear in $B$, as many previous model calculations showed, but also that they have 
a nonzero intercept (offset), which appears as a result of the Landau level mixing.

\begin{figure*}
\begin{center}
\includegraphics[width=.4\textwidth]{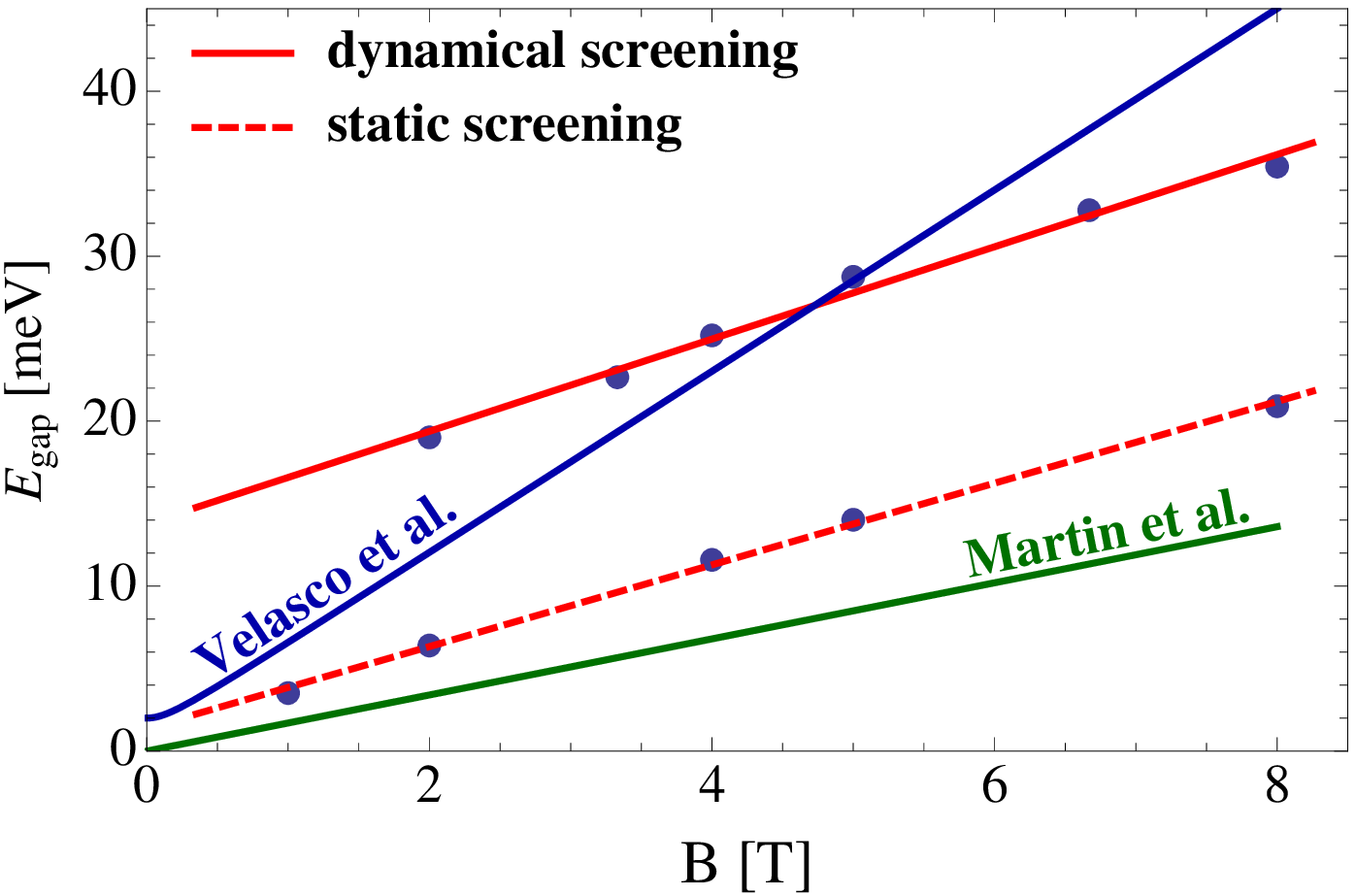}\hspace{0.08\textwidth}
\includegraphics[width=.4\textwidth]{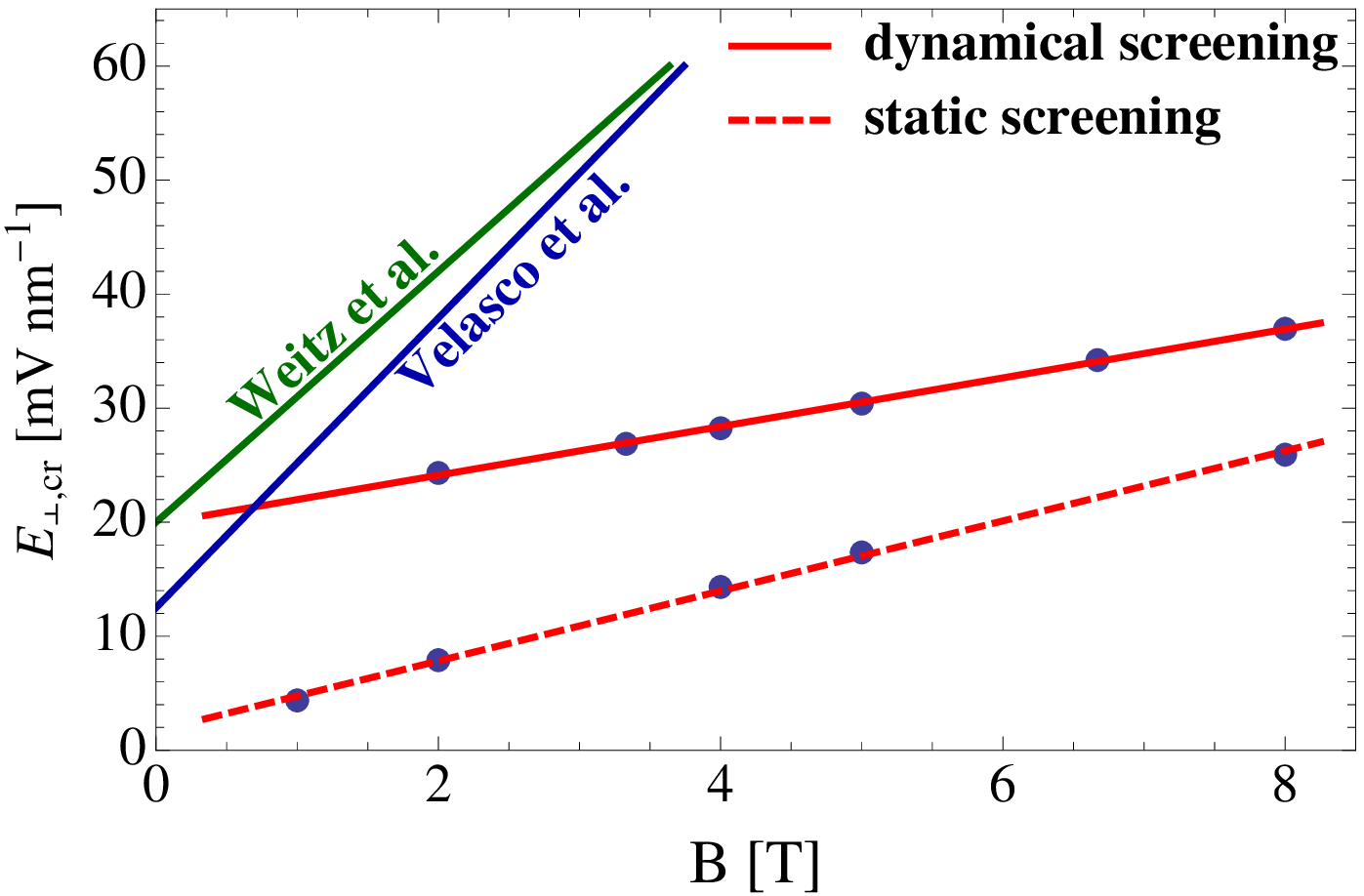}
\caption{(Color online) Comparison of the energy gap $E_{\rm gap}$ in the SP state as a function of the 
magnetic field at $E_{\perp}= 0$ with the experimental results of Refs.~\onlinecite{Martin,Velasco} (left panel). 
Comparison of the dependence of the critical value of $E_{\perp}$ on the magnetic field with the experimental 
results of Refs.~\onlinecite{Weitz,Velasco} (right panel).}
\label{E-gap-vs-B}
\end{center}
\end{figure*}

The results for the SP gap (at $\tilde{\Delta}_{0}=0$) as a function of the magnetic field are 
shown in the left panel in Fig.~\ref{E-gap-vs-B}. As we see, the results in the case of static screening
are in a reasonable agreement with the experiment in the low-mobility graphene samples.\cite{Martin}
The results for the gaps in the case of dynamical screening are in order of magnitude agreement with 
the experiment in the high-mobility graphene samples.\cite{Velasco} On the other hand, the
slopes and the intercepts of theoretical lines are not in excellent quantitative agreement.
There may exist many potential reasons for the discrepancies. One of them is the limitations
of the model itself, for example, the use of the zero width of the Landau levels. The other is the 
approximations used in the analysis of the model. Perhaps the most important limitation of the second 
type is the energy independent ansatz for the gap parameters.

As to the dependence of the critical electric field $E_{\perp,{\rm cr}}$ on the magnetic
field in Eq.~(\ref{linear-offset}), we find that it is a linear function of $B$ just like in the 
experiment\cite{Weitz,Velasco} and that the value of the offset is in a reasonable
agreement with the experimental values (see the right panel in Fig.~\ref{E-gap-vs-B}).

The theoretical slope $2.13\,\mbox{mV}\, \mbox{nm}^{-1}\,\mbox{T}^{-1}$, however, appears to be much 
smaller than the experiment value, $12.7 \,\mbox{mV}\, \mbox{nm}^{-1}\,\mbox{T}^{-1}$. This discrepancy may have 
its roots in disorder, which presumably plays a profound role in real samples. One may suggest, for example,
that an external electric field is more effective in clean samples (considered in our model) and,
therefore, its critical value $E_{\perp,{\rm cr}}$ should be smaller than that in samples with impurities. This 
conjecture agrees with that $E_{\perp,{\rm cr}}$ is smaller in the experiment\cite{Velasco} with high-mobility samples
than that in the experiment\cite{Weitz} with low-mobility ones (see the right panel in Fig.~\ref{E-gap-vs-B} above).

It is appropriate to mention another key experimental observation in Ref.~\onlinecite{Weitz}
that the two-terminal conductance is quantized except at particular values of the electric field $E_\perp$. 
For the $\nu =0$ state, the experimental data shows that the conductance is quantized except for two 
values of $E_\perp$.\cite{Weitz} Let us argue that these two values correspond 
to $E_\perp = \pm E_{\perp,{\rm cr}}$, where $E_{\perp,{\rm cr}}$ is the critical value at which
the phase transition between the SP and LP phases takes place (see Fig.~\ref{fig-E-gap}, where
we consider only nonnegative values of $\tilde{\Delta}_0=e E_{\perp}d /2$).
As one can see in this figure, the gap $\Delta_{\nu=0}$ has a maximum at $E_{\perp}=0$ and a
minimum at $E_{\perp}=E_{\perp,{\rm cr}}$. This implies that while the conductivity is suppressed 
almost everywhere, it is enhanced for the two values of $E_{\perp}=\pm E_{\perp,{\rm cr}}$ 
(compare with Fig.~2C in Ref.~\onlinecite{Weitz}).

In this paper, we did not investigate the limit of weak magnetic fields, which is outside
the scope of the present paper. In fact, the numerical
approach used here is not efficient in such a limit because the number of Landau
levels becomes too large and the importance of level discretization itself diminishes.
It is still interesting to speculate about the physical meaning of our result when
extrapolated to vanishing fields. A non-zero value of the offset electric field
$E_{\perp}^{\rm off}$, needed in the fit of the critical line, suggests that the ground
state of bilayer graphene without magnetic and electric fields
can be a spin-polarized (ferromagnetic) state. In the future
studies, it will be interesting to address this question in detail by considering the weak
field limit.

In the recent experimental paper,\cite{Mayorov} a phase transition to the nematic state was observed
in bilayer graphene without magnetic field. The nematic state has been predicted in theoretical
works.\cite{Vafek,Lemonik} It breaks the rotational symmetry and keeps quasiparticles
gapless.\cite{Cserti} It would
be interesting to study a competition between the gapped state and the nematic state due to
the Coulomb interaction and other interactions in bilayer graphene in a magnetic field.

In the future, it would be also interesting to extend our study to the case of a more general
model of bilayer graphene formulated in terms of the four-component spinors, which
has a much larger energy range of applicability and allows to address the competition
of the SP and LP states in the very strong magnetic fields. The dynamics at larger
energies should reveal an interesting cross-over to the regime when the top and
bottom layers of bilayer effectively decouple. 

It will be also interesting to extend the analysis of this paper to the case of 
quantum Hall states with nonzero filling factors. It is natural to expect that the Landau level
mixing and dynamical screening will again substantially modify the corresponding results 
obtained in the LLL approximation.\cite{Barlas,Nandkishore1,GGM1,GGM2,Nandkishore2,Toke,Kharitonov1,GGJM} 
Finally, there remains the question whether there are nonuniform 
states, e.g., such as the helical and electron crystal states,\cite{Cote} that are energetically 
more favorable in bilayer graphene under certain conditions.

\begin{acknowledgments}
The work of E.V.G and V.P.G. was supported partially by the Scientific Cooperation
Between Eastern Europe and Switzerland (SCOPES) program under Grant
No.~IZ73Z0-128026 of the Swiss National Science Foundation (NSF),  the European
FP7 program, Grant No. SIMTECH 246937, the joint Ukrainian-Russian SFFR-RFBR 
Grant No.~F40.2/108, and by the Program of Fundamental Research of the Physics 
and Astronomy Division of the NAS of Ukraine. V.P.G. acknowledges a collaborative 
grant from the Swedish Institute. The work of V.A.M. was supported by the Natural Sciences 
and Engineering Research Council of Canada. The work of I.A.S. is supported in part by
the U.S. National Science Foundation under Grant No. PHY-0969844.
\end{acknowledgments}

\appendix


\begin{widetext}

\section{Quasiparticle Green's function}
\label{Quasiparticle propagator}

The quasiparticle Green's function can be obtained in a standard way by making use of a complete
set of eigenstates for the low-energy Hamiltonian in bilayer graphene in a magnetic field.
Schematically,
\begin{equation}
S(\omega;\bm{r},\bm{r}^\prime)  = \sum_{\{k_i\}}\frac{\Psi_{\{k_i\}}(\bm{r})\otimes
\Psi_{\{k_i\}}^{\dagger}(\bm{r}^\prime)}{\omega-E_{\{k_i\}}},
\end{equation}
where $\{k_i\}$ stands for a complete set of quantum numbers that uniquely define the
eigenstates. Here and in the rest of appendixes we put $\hbar=1$.

In the Landau gauge, $\bm{A}= (A_x, A_y, A_z) = (0, B_{\perp}x, 0)$, the corresponding set
of eigenfunctions for the free Hamiltonian reads
\begin{eqnarray}
\Psi_{k,\xi,n,\tau=-1}(\bm{r})&=& e^{iky} \left(\begin{array}{l}
0 \\
u_{n}(\chi)
\end{array}\right),\qquad n=0,1 ,\\
\Psi_{k,\xi,n,\tau}(\bm{r})&=&\frac{e^{iky}}{\sqrt{2M_{n}}}\left(\begin{array}{l}
\sqrt{M_{n}+\tau\xi \tilde{\Delta}_{0}} \,u_{n-2}(\chi) \\
\tau\sqrt{M_{n}-\tau\xi \tilde{\Delta}_{0}}\, u_{n}(\chi)
\end{array}\right),\qquad n\geq 2 ,
\end{eqnarray}
where $\chi=k\ell+x/\ell$,
$M_{n}=\sqrt{\tilde{\Delta}_{0}^2+\omega_{c}^2n(n-1)}$, $\tau=\pm$,
and
\begin{equation}
u_{n}(\chi)  = \frac{1}{\sqrt{2\pi\ell}}\frac{e^{-\chi^2/2}}{\sqrt{2^n n! \sqrt{\pi}}} H_{n}(\chi).
\end{equation}
The corresponding energy eigenvalues are $\pm M_{n}$.

By making use of the above complete set of eigenstates, it is straightforward to derive the
following explicit form of the free quasiparticle Green's function:
\begin{eqnarray}
S(\omega;\bm{r},\bm{r}^\prime) &=& e^{i\Phi(\bm{r},\bm{r}^\prime)}\tilde{S}(\omega;\bm{r}-\bm{r}^\prime) ,\\
\tilde{S}(\omega;\bm{r})&=& \frac{e^{-z/2}}{2\pi\ell^2}
\Bigg\{
\frac{L_{0}(z)+L_{1}(z)}{\omega+\bar{\mu}_{s}+\xi \tilde{\Delta}_{0}}{\cal P}_{-}
+\sum_{n=2}^{n_{\rm max}}\left[\frac{(\omega+\bar{\mu}_{s}-\xi \tilde{\Delta}_{0})
L_{n}(z)}{(\omega+\bar{\mu}_{s})^2-M_{n}^{2}}{\cal P}_{-}
+\frac{(\omega+\bar{\mu}_{s}+\xi \tilde{\Delta}_{0})L_{n-2}(z)}{(\omega+\bar{\mu}_{s})^2
-M_{n}^{2}}{\cal P}_{+}\right] \nonumber \\
&&+\sum_{n=2}^{n_{\rm max}}\frac{\omega_{c} }{2[(\omega+\bar{\mu}_{s})^2-M_{n}^{2}]\ell^2}
\left(\begin{array}{ll}
0 & r_{-}^2 \\
r_{+}^2 & 0
\end{array}\right)L^{2}_{n-2}(z)
\Bigg\},
\end{eqnarray}
where $z=\bm{r}^2/(2\ell^2)$,
$\bar{\mu}_{s}=\mu_{0}-s Z$,
$r_{\pm} = x\pm i y$ and ${\cal P}_{\pm} =(1\pm\tau_{3})/2$.
Note that the translational invariance of this function is spoiled only by the Schwinger phase
factor $e^{i\Phi(\bm{r},\bm{r}^\prime)}$. In the Landau gauge used here, the explicit
form of the phase reads
\begin{equation}
\Phi(\bm{r},\bm{r}^\prime)=-\frac{(x+x')(y-y')}{2\ell^{2}}{\mbox{sign}(eB_\perp).}
\end{equation}

Using the same approach, we can also derive a similar representation for the inverse Green's function:
\begin{eqnarray}
S^{-1}(\omega;\bm{r},\bm{r}^\prime) &=& e^{i\Phi(\bm{r},\bm{r}^\prime)}\tilde{S}^{-1}(\omega;\bm{r}-\bm{r}^\prime) ,\\
\tilde{S}^{-1}(\omega;\bm{r})&=& \frac{e^{-z/2}}{2\pi\ell^2}
\Bigg\{
(\omega+\bar{\mu}_{s}+\xi \tilde{\Delta}_{0})\left[L_{0}(z)+L_{1}(z)\right]{\cal P}_{-}\nonumber\\
&+&\sum_{n=2}^{n_{\rm max}}\left[(\omega+\bar{\mu}_{s}+\xi \tilde{\Delta}_{0})L_{n}(z){\cal P}_{-}
+(\omega+\bar{\mu}_{s}-\xi \tilde{\Delta}_{0})L_{n-2}(z){\cal P}_{+}
- \frac{\omega_{c} }{2\ell^2}\left(\begin{array}{ll}
0 & r_{-}^2 \\
r_{+}^2 & 0
\end{array}\right)L^{2}_{n-2}(z) \right]
\Bigg\}.
\end{eqnarray}
Let us emphasize that the Green's function and its inverse have the same Schwinger phases.
One can also show that the phase remains exactly the same even for the full quasiparticle Green's function
(see below). This property plays an important role in the derivation of the gap (Schwinger-Dyson) equation,
which takes a very simple form after the same overall non-zero Schwinger phase on both sides of the
equation is eliminated.

The full Green's function, which incorporates the effects due to dynamically generated order parameters
of quantum Hall ferromagnetism
and magnetic catalysis, can be derived in the same way as
the free function. The final result reads
\begin{eqnarray}
\label{A9}
G(\omega;\bm{r},\bm{r}^\prime) &=& e^{i\Phi(\bm{r},\bm{r}^\prime)}\tilde{G}(\omega;\bm{r}-\bm{r}^\prime) ,\\
\label{A10}
\tilde{G}(\omega;\bm{r})&=& \frac{e^{-z/2}}{2\pi\ell^2}
\Bigg\{
\frac{L_{0}(z)}{\omega-E_{\xi 0 s}^{L} }{\cal P}_{-}
+\frac{L_{1}(z)}{\omega-E_{\xi 1 s}^{L} }{\cal P}_{-}
+\sum_{n=2}^{n_{\rm max}}\frac{\omega_{c} }{2[(\omega+\mu_{\xi ns})^2-M_{\xi ns}^{2}]\ell^2}\left(\begin{array}{ll}
0 & r_{-}^2 \\
r_{+}^2 & 0
\end{array}\right)L^{2}_{n-2}(z)
\nonumber\\
&+&\sum_{n=2}^{n_{\rm max}}\left[\frac{(\omega+\mu_{\xi ns}-\Delta_{\xi ns})L_{n}(z)}{(\omega+\mu_{\xi ns})^2-M_{\xi ns}^{2}}{\cal P}_{-}
+\frac{(\omega+\mu_{\xi ns}+\Delta_{\xi ns})L_{n-2}(z)}{(\omega+\mu_{\xi ns})^2-M_{\xi ns}^{2}}{\cal P}_{+}\right]
\Bigg\},
\end{eqnarray}
where
{\begin{eqnarray}
\mu_{\xi ns}&=& \mu_{ns}-\xi\tilde{\mu}_{ns},\\
\Delta_{\xi ns}&=& \xi \tilde{\Delta}_{ns}+\Delta_{ns},\\
\label{nLLL}
E_{\xi n s}^{L} &=& - \Delta_{ns}  - \xi \tilde{\Delta}_{ns},\quad \mbox{for} \quad n=0,1,\\
M_{\xi ns}&=&\sqrt{\Delta_{\xi ns}^2+\omega_{c}^2n(n-1)},\quad \mbox{for} \quad n\geq 2.
\end{eqnarray}}
The corresponding inverse Green's function is
\begin{eqnarray}
\label{A15}
G^{-1}(\omega;\bm{r},\bm{r}^\prime) &=& e^{i\Phi(\bm{r},\bm{r}^\prime)}\tilde{G}^{-1}(\omega;\bm{r}-\bm{r}^\prime) ,\\
\label{A16}
\tilde{G}^{-1}(\omega;\bm{r})&=& \frac{e^{-z/2}}{2\pi\ell^2}
\Bigg\{
(\omega-E_{\xi 0 s}^{L} ) L_{0}(z) {\cal P}_{-}
+(\omega-E_{\xi 1 s}^{L} ) L_{1}(z) {\cal P}_{-}
-\sum_{n=2}^{n_{\rm max}} \frac{\omega_{c} }{2\ell^2}\left(\begin{array}{ll}
0 & r_{-}^2 \\
r_{+}^2 & 0
\end{array}\right)L^{2}_{n-2}(z)
\nonumber\\
&+& \sum_{n=2}^{n_{\rm max}}\left[(\omega+\mu_{\xi ns}+\Delta_{\xi ns})L_{n}(z){\cal P}_{-}
+(\omega+\mu_{\xi ns}-\Delta_{\xi ns})L_{n-2}(z){\cal P}_{+} \right]
\Bigg\}.
\end{eqnarray}
While $E_{\xi n s}^{L} $ are the quasiparticle energies in the two lowest Landau levels, the quasiparticle energies
in higher Landau levels ($n\geq 2$) are given by the following expression:
\begin{eqnarray}
E_{\xi n s} &=& -\mu_{\xi ns}\pm M_{\xi ns},\quad \mbox{for} \quad n\geq 2.
\label{n2}
\end{eqnarray}
It should be emphasized that, while the Green's functions $G(\omega;\bm{r},\bm{r}^\prime)$ and
$G^{-1}(\omega;\bm{r},\bm{r}^\prime)$ are inverse of each other, their translationally invariant parts,
$\tilde{G}(\omega;\bm{r})$ and $\tilde{G}^{-1}(\omega;\bm{r})$, are not. Another important property,
which is used in the derivation of the gap equation, is that the Schwinger phases of the quasiparticle
Green's function and its inverse are identical.

\section{Interaction coefficient functions $I_{n^\prime,n}(E)$}
\label{Tables}

The interaction coefficient functions $I_{n^\prime,n}(E)$ entering the gap equations
(\ref{dyn-gap-eq-n=0}) -- (\ref{dyn-gap-eq-n>1B}) are given by the following expression,
\begin{equation}
I_{n^{\prime},n}(E) = \int_{0}^{\infty}
\frac{2 {\cal L}_{n^{\prime},n} (y)dy}{\pi \kappa\sqrt{xy}(\bar{E}_y^2-1+\Sigma_y^2)}\left[
\frac{\Sigma_y^2\bar{E}_y}{\sqrt{1 -\Sigma_y^2 }}\arctan\frac{\sqrt{1 -\Sigma_y^2 }}{\Sigma_y}
+\frac{\pi}{2}(\bar{E}_y^2-1) +\Sigma_y\sqrt{1-\bar{E}_y^2}\arctan\frac{\sqrt{1-\bar{E}_y^2}}{\bar{E}_y}
\right],
\label{Innprime}
\end{equation}
where functions ${\cal L}_{n^{\prime},n} (y)$ are defined in Eq.~(\ref{L_nm}) and
\begin{eqnarray}
\bar{E}_y &=& \frac{|E|}{\hbar\omega_c}\sqrt{\frac{b_4}{1 + b_2 y + b_3 y^2}},\\
\Sigma_y &=& \frac{4\pi(1+b_1 y)\tilde{\Pi}(y)}{\kappa\sqrt{xy}\sqrt{1 + b_2 y + b_3 y^2}}.
\end{eqnarray}
In the case of model interaction with a static screening the corresponding coefficient functions
$I_{n^\prime,n}$ are energy independent. They can be easily obtained from the more
general dynamical expression in Eq.~(\ref{Innprime}) by substituting $E=0$ and taking $b_i=0$ ($i=1,4$),
i.e.,
\begin{equation}
 I_{n,n^\prime} = \int_{0}^{\infty}\frac{{\cal L}_{n,n^\prime} (y)dy}{\kappa\sqrt{xy} +4\pi\tilde{\Pi}(y)} .
\end{equation}
Now, in the case of a dynamical screening, the interaction coefficient functions
$I_{n^\prime,n}(E)$ become energy dependent. Using the definition in Eq.~(\ref{Innprime}), we can
easily tabulate these functions. The resulting dependence can be fitted by Pad\'{e} approximants of
order [2/1]:
\begin{equation}
I_{n^{\prime},n}(E) = \frac{a_{n^{\prime},n}+ b_{n^{\prime},n} \frac{|E|}{\hbar\omega_c} + c_{n^{\prime},n}
\left(\frac{E}{\hbar\omega_c} \right)^2}{1+d_{n^{\prime},n} \frac{|E|}{\hbar\omega_c} } .
\label{Innprime-Pade-app}
\end{equation}
Such fits are within a few percent of the numerical results. For the functions $I_{n^\prime,n}(E)$ with
not very large indices, $n^\prime,n\leq 10$, the fits were obtained from the data in the energy range
$0<E\lesssim 12 \hbar\omega_c$, while for larger indices, $10\leq n^\prime,n\leq 40$, we were also
using data in a wider range of energies,  $0<E\lesssim 40 \hbar\omega_c$. A subset of the corresponding
fit parameters for $I_{n^\prime,n}(E)$ with $0\leq n^{\prime},n\leq 10$ in the case of $B=2\,\mbox{T}$
and $\kappa=2$ is presented in Table~\ref{tab-a-b-c-d}. Note that we show only the upper triangular
part of the corresponding tables because all coefficients are symmetric with respect to the exchange
$n^{\prime}$ and $n$.

\begin{table}
\caption{Coefficients $a_{n,n^\prime}$, $b_{n,n^\prime}$, $c_{n,n^\prime}$ and $d_{n,n^\prime}$,
evaluated for $B=2~\mbox{T}$ and $\kappa=2$.}
\begin{ruledtabular}
\begin{tabular}{c|ccccccccccc}
$a_{n,n^\prime}$    &   $n^\prime=0$ & $n^\prime=1$ & $n^\prime=2$ & $n^\prime=3$ & $n^\prime=4$ &
$n^\prime=5$ & $n^\prime=6$ & $n^\prime=7$ & $n^\prime=8$ & $n^\prime=9$ & $n^\prime=10$ \\
 \hline
     $n=0$ & 1.2315 & 0.2959 & 0.2215 & 0.1943 & 0.1806 & 0.1733 & 0.1694 & 0.1674 & 0.1664 & 0.166 & 0.1659 \\
     $n=1$ &      & 1.0827 & 0.2888 & 0.2214 & 0.1987 & 0.1866 & 0.1789 & 0.1737 & 0.1703 & 0.1681 & 0.1667 \\
     $n=2$ &      &      & 1.0151 & 0.2905 & 0.2231 & 0.2004 & 0.1887 & 0.1813 & 0.176 & 0.1722 & 0.1694 \\
     $n=3$ &      &      &      & 0.9712 & 0.2915 & 0.2241 & 0.2012 & 0.1896 & 0.1823 & 0.1771 & 0.1731 \\
     $n=4$ &      &      &      &      & 0.9381 & 0.2918 & 0.2246 & 0.2016 & 0.1899 & 0.1826 & 0.1775 \\
     $n=5$ &      &      &      &      &      & 0.9116 & 0.2917 & 0.2247 & 0.2016 & 0.1898 & 0.1826 \\
     $n=6$ &      &      &      &      &      &      & 0.8893 & 0.2913 & 0.2245 & 0.2014 & 0.1896 \\
     $n=7$ &      &      &      &      &      &      &      & 0.87 & 0.2907 & 0.2243 & 0.2011 \\
     $n=8$ &      &      &      &      &      &      &      &      & 0.8532 & 0.2901 & 0.2239 \\
     $n=9$ &      &      &      &      &      &      &      &      &      & 0.8382 & 0.2894 \\
     $n=10$ &      &      &      &      &      &      &      &      &      &      & 0.8247
\end{tabular}
\end{ruledtabular}
\begin{ruledtabular}
\begin{tabular}{c|ccccccccccc}
$b_{n,n^\prime}$  &  $n^\prime=0$ & $n^\prime=1$ & $n^\prime=2$ & $n^\prime=3$ & $n^\prime=4$ &
$n^\prime=5$ & $n^\prime=6$ & $n^\prime=7$ & $n^\prime=8$ & $n^\prime=9$ & $n^\prime=10$ \\
  \hline
    $n=0$ & 1.186 & 0.2911 & 0.1683 & 0.1186 & 0.0915 & 0.0745 & 0.0631 & 0.0548 & 0.0486 & 0.0437 & 0.0396 \\
     $n=1$ &      & 0.957 & 0.2684 & 0.1617 & 0.1175 & 0.0923 & 0.0757 & 0.064 & 0.0555 & 0.049 & 0.044 \\
     $n=2$ &      &      & 0.8567 & 0.257 & 0.1572 & 0.1157 & 0.0919 & 0.0761 & 0.0647 & 0.0561 & 0.0496 \\
     $n=3$ &      &      &      & 0.794 & 0.2491 & 0.1538 & 0.114 & 0.0913 & 0.0761 & 0.0651 & 0.0567 \\
     $n=4$ &      &      &      &      & 0.749 & 0.2429 & 0.1512 & 0.1126 & 0.0906 & 0.076 & 0.0652 \\
     $n=5$ &      &      &      &      &      & 0.7143 & 0.2379 & 0.149 & 0.1114 & 0.09 & 0.0757 \\
     $n=6$ &      &      &      &      &      &      & 0.6862 & 0.2337 & 0.1471 & 0.1103 & 0.0894 \\
     $n=7$ &      &      &      &      &      &      &      & 0.6627 & 0.23 & 0.1455 & 0.1094 \\
     $n=8$ &      &      &      &      &      &      &      &      & 0.6425 & 0.2268 & 0.144 \\
     $n=9$ &      &      &      &      &      &      &      &      &      & 0.6249 & 0.2239 \\
     $n=10$ &      &      &      &      &      &      &      &      &      &      & 0.6093
\end{tabular}
\end{ruledtabular}
\begin{ruledtabular}
\begin{tabular}{c|ccccccccccc}
$c_{n,n^\prime}$ & $n^\prime=0$ & $n^\prime=1$ & $n^\prime=2$ & $n^\prime=3$ & $n^\prime=4$ &
$n^\prime=5$ & $n^\prime=6$ & $n^\prime=7$ & $n^\prime=8$ & $n^\prime=9$ & $n^\prime=10$ \\
 \hline
     $n=0$ & 0.0177 & 0.0049 & 0.0027 & 0.0018 & 0.0013 & 0.001 & 0.0008 & 0.0006 & 0.0005 & 0.0004 & 0.0004 \\
     $n=1$ &      & 0.0138 & 0.0047 & 0.0027 & 0.0019 & 0.0014 & 0.0011 & 0.0008 & 0.0007 & 0.0006 & 0.0005 \\
     $n=2$ &      &      & 0.012 & 0.0046 & 0.0027 & 0.0019 & 0.0014 & 0.0011 & 0.0009 & 0.0007 & 0.0006 \\
     $n=3$ &      &      &      & 0.0109 & 0.0044 & 0.0027 & 0.0019 & 0.0015 & 0.0012 & 0.0009 & 0.0008 \\
     $n=4$ &      &      &      &      & 0.0101 & 0.0043 & 0.0027 & 0.0019 & 0.0015 & 0.0012 & 0.001 \\
     $n=5$ &      &      &      &      &      & 0.0095 & 0.0042 & 0.0027 & 0.0019 & 0.0015 & 0.0012 \\
     $n=6$ &      &      &      &      &      &      & 0.0089 & 0.0041 & 0.0026 & 0.0019 & 0.0015 \\
     $n=7$ &      &      &      &      &      &      &      & 0.0085 & 0.004 & 0.0026 & 0.0019 \\
     $n=8$ &      &      &      &      &      &      &      &      & 0.0082 & 0.0039 & 0.0026 \\
     $n=9$ &      &      &      &      &      &      &      &      &      & 0.0079 & 0.0038 \\
     $n=10$ &      &      &      &      &      &      &      &      &      &      & 0.0076
\end{tabular}
\end{ruledtabular}
\begin{ruledtabular}
\begin{tabular}{c|ccccccccccc}
$d_{n,n^\prime}$ & $n^\prime=0$ & $n^\prime=1$ & $n^\prime=2$ & $n^\prime=3$ & $n^\prime=4$ &
$n^\prime=5$ & $n^\prime=6$ & $n^\prime=7$ & $n^\prime=8$ & $n^\prime=9$ & $n^\prime=10$ \\
 \hline
    $n=0$ & 0.2047 & 0.1273 & 0.1041 & 0.0902 & 0.08 & 0.0721 & 0.0659 & 0.0608 & 0.0565 & 0.0529 & 0.0497 \\
     $n=1$ &      & 0.2207 & 0.1373 & 0.1118 & 0.0973 & 0.0867 & 0.0784 & 0.0715 & 0.0659 & 0.0611 & 0.0571 \\
     $n=2$ &      &      & 0.2305 & 0.1446 & 0.1176 & 0.1027 & 0.0921 & 0.0835 & 0.0765 & 0.0705 & 0.0654 \\
     $n=3$ &      &      &      & 0.2376 & 0.1506 & 0.1224 & 0.1071 & 0.0964 & 0.0878 & 0.0807 & 0.0745 \\
     $n=4$ &      &      &      &      & 0.2433 & 0.1556 & 0.1265 & 0.1109 & 0.1 & 0.0915 & 0.0843 \\
     $n=5$ &      &      &      &      &      & 0.248 & 0.16 & 0.1301 & 0.1142 & 0.1032 & 0.0946 \\
     $n=6$ &      &      &      &      &      &      & 0.2519 & 0.1639 & 0.1333 & 0.1171 & 0.106 \\
     $n=7$ &      &      &      &      &      &      &      & 0.2554 & 0.1673 & 0.1362 & 0.1197 \\
     $n=8$ &      &      &      &      &      &      &      &      & 0.2583 & 0.1703 & 0.1388 \\
     $n=9$ &      &      &      &      &      &      &      &      &      & 0.261 & 0.1731 \\
     $n=10$ &      &      &      &      &      &      &      &      &      &      & 0.2634

\end{tabular}
\end{ruledtabular}
\label{tab-a-b-c-d}
\end{table}

\section{Free energy density}
\label{Free energy}

The energy of a given solution is defined by the value of the Baym--Kadanoff effective action
calculated on this solution. In our analysis, we use the two-loop effective action
given in Ref.~\onlinecite{GGJM}. Then for the energy of the system, we have
\begin{eqnarray}
V &=& i\int\limits_{-\infty}^{\infty}\frac{d\omega}{2\pi} \mbox{Tr}
\left\{-\omega\frac{\partial G^{-1}(\omega)} {\partial\omega}\,G(\omega)
+\frac{1}{2}\left[ S^{-1}(\omega) \,G(\omega)-1\right]\right\}.
\label{effpot2}
\end{eqnarray}
By making use of the relation
\begin{equation}
\frac{\partial G^{-1}(\omega;\mathbf{r},\mathbf{r}^\prime)}{\partial\omega} =
\delta(\mathbf{r}-\mathbf{r}^\prime),
\end{equation}
we obtain the energy expressed through the translation invariant part of quasiparticle Green's function
\begin{eqnarray}
V &=& i\Omega \int\limits_{-\infty}^{\infty}\frac{d\omega}
{2\pi}\mbox{tr} \left[-\omega \, \tilde{G}(\omega;\mathbf{0})
+\frac{1}{2}\int d^2 \mathbf{r} \, \tilde{S}^{-1}(\omega;\mathbf{r}) \,\tilde{G}(\omega;-\mathbf{r}) \right]
-V_0\,,
\end{eqnarray}
where the overall factor $\Omega$ is the space volume. Dividing $V$ by this volume and performing all
integrations and traces, we derive the following expression for the energy density of the system:
\begin{eqnarray}
{\cal V} &=& -\frac{i}{4\pi\ell^2} \int\limits_{-\infty}^{\infty}\frac{d\omega}{2\pi}
\sum_{\xi,s}  \Bigg[
\frac{\omega-\bar{\mu}_{s}-\xi \tilde{\Delta}_{0}}{\omega-E^{L}_{\xi 0 s}}
+\frac{\omega-\bar{\mu}_{s}-\xi \tilde{\Delta}_{0}}{\omega-E^{L}_{\xi 1 s}} +2\sum_{n=2}^{n_{\rm max}}
\frac{(\omega- \bar{\mu}_{s})(\omega+\mu_{\xi n s})+\xi \tilde{\Delta}_{0}\Delta_{\xi n s}
+\omega_{c}^{2}n(n-1)}{(\omega+\mu_{\xi n s})^2-M_{\xi n s}^2}
\Bigg]-{\cal V}_{0}\nonumber\\
&=&-\frac{1}{4\pi \ell^2}\sum_{\xi,s}  \Bigg\{
\frac{1}{2}\left(E^{L}_{\xi 0 s}-\bar{\mu}_{s}-\xi \tilde{\Delta}_{0}\right)\mbox{sign}
\left(E^{L}_{\xi 0 s}\right)+\frac{1}{2}\left(E^{L}_{\xi 1 s}-\bar{\mu}_{s}-
\xi \tilde{\Delta}_{0}\right)\mbox{sign}\left(E^{L}_{\xi 1 s}\right)\nonumber\\
&&+\sum_{n=2}^{n_{\rm max}} \left[
(\bar{\mu}_{s}+\mu_{\xi n s})\mbox{sign}\left(\mu_{\xi n s}\right)\theta\left(\mu_{\xi n s}^2-M_{\xi n s}^2\right)
+\frac{\Delta_{\xi n s}(\Delta_{\xi n s}+\xi \tilde{\Delta}_{0})+2\omega_{c}^{2}n(n-1)}{M_{\xi n s}}
\theta\left(M_{\xi n s}^2-\mu_{\xi n s}^2\right)
\right]\nonumber\\
&&-\sum_{n=2}^{n_{\rm max}} 2\omega_{c}\sqrt{n(n-1)}
\Bigg\}.
\label{V-energy}
\end{eqnarray}
Note that the sum over Landau levels in the last expression has the following asymptotic behavior
at large $n$:
\begin{eqnarray}
-\frac{1}{4\pi \ell^2}\frac{\tilde{\Delta}_{0}}{\omega_{c} }
\sum_{\xi,s} \sum_{n=2}^{n_{\rm max}}\frac{\xi \Delta_{\xi n s}}{n}\,,
\end{eqnarray}
which is convergent only if the layer- and spin-averaged gap decreases fast enough with $n$.

In the framework of the canonical ensemble, the energy of a system is replaced by its free energy through
the corresponding Legendre transformation. For the system under consideration, we find the following free
energy density:
\begin{equation}
{\cal F}= {\cal V} +\mu_{0} \rho,
\end{equation}
where
\begin{equation}
\rho=i\int_{-\infty}^{\infty}\frac{d\omega^{\prime}}{2\pi}\mbox{tr}\left[\tilde{G}(\omega^{\prime};0)\right]
=-\frac{1}{4\pi\ell^2} \sum_{\xi,s} \left[
\mbox{sign}(E_{\xi 0 s}^{L} ) +\mbox{sign}(E_{\xi 1 s}^{L} ) -2 \sum_{n=2}^{n_{\rm max}}
 \mbox{sign}(\mu_{\xi ns})\theta(\mu_{\xi ns}^2-M_{\xi ns}^{2})
\right].
\end{equation}
Finally, by making use of the explicit expression for the energy density in Eq.~(\ref{V-energy}), we obtain
\begin{eqnarray}
{\cal F} &=&-\frac{1}{4\pi \ell^2}\sum_{\xi,s}  \Bigg\{
\frac{1}{2}\left(E^{L}_{\xi 0 s}+\mu_{0}+sZ-\xi \tilde{\Delta}_{0}\right)\mbox{sign}\left(E^{L}_{\xi 0 s}\right)
+\frac{1}{2}\left(E^{L}_{\xi 1 s}+\mu_{0}+sZ-\xi \tilde{\Delta}_{0}\right)\mbox{sign}\left(E^{L}_{\xi 1 s}\right)
\nonumber \\
&&+\sum_{n=2}^{n_{\rm max}} \left[
(\mu_{\xi n s}-\mu_{0}-sZ)\mbox{sign}\left(\mu_{\xi n s}\right)\theta\left(\mu_{\xi n s}^2-M_{\xi n s}^2\right)
+\frac{\Delta_{\xi n s}(\xi \tilde{\Delta}_{0}-\Delta_{\xi n s})+2\omega_{c}^{2}M_{\xi n s}}{M_{\xi n s}}\,
\theta\left(M_{\xi n s}^2-\mu_{\xi n s}^2\right)
\right]\nonumber\\
&&-\sum_{n=2}^{n_{\rm max}} 2\omega_{c}\sqrt{n(n-1)}
\Bigg\}.
\label{B8}
\end{eqnarray}

\end{widetext}


\begin{thebibliography}{99}

\bibitem{Falko}
E.~McCann and V.I.~Fal'ko, Phys. Rev. Lett. {\bf 96}, 086805 (2006).

\bibitem{Ohta}
T.~Ohta, A.~Bostwick, T.~Seyller, K.~Horn, and E.~Rotenberg, Science {\bf 313}, 951 (2006).

\bibitem{Zhang}
R.~Nandkishore, L.~Levitov, Phys. Rev. Lett. {\bf 104}, 156803 (2010);
F.~Zhang, H.~Min, M.~Polini, A.H.~MacDonald, Phys. Rev. B {\bf 81}, 041402 (2010).

\bibitem{Sun}
K.~Sun and E.~Fradkin, Phys. Rev. B {\bf 78}, 245122 (2008);
K.~Sun, H.~Yao, E.~Fradkin, and S.A.~Kivelson, Phys. Rev. Lett. {\bf 103}, 046811 (2009).

\bibitem{Levitov-anomalous}
R.~Nandkishore and L.~Levitov, Phys. Rev. B {\bf 82}, 115124 (2010).

\bibitem{Fzhang}
F.~Zhang, J.~Jung, G.~A.~Fiete, Q.~Niu, and A.~H.~MacDonald, Phys. Rev. Lett. {\bf 106}, 156801 (2011).

\bibitem{Vafek}
O.~Vafek and K.~Yang, Phys. Rev. B {\bf 81}, 041401 (2010).

\bibitem{Lemonik}
Y.~Lemonik, I.L.~Aleiner, C.~T\H{o}ke, and V.I.~Fal'ko, Phys. Rev. B {\bf 82}, 201408 (2010).

\bibitem{MacDonald}
F.~Zhang and A.H.~MacDonald, Phys. Rev. Lett. {\bf 108}, 186804 (2012).

\bibitem{catal}
V.~P.~Gusynin, V.~A.~Miransky, and I.~A.~Shovkovy,
Phys. Rev. Lett. {\bf 73}, 3499 (1994);
Phys. Rev.  D {\bf 52}, 4718 (1995).

\bibitem{Khvesh}
D.~V.~Khveshchenko, Phys. Rev. Lett. {\bf 87}, 206401 (2001).

\bibitem{graphite}
E.V.~Gorbar, V.P.~Gusynin, V.A.~Miransky, and I.A.~Shovkovy,
Phys. Rev. B {\bf 66}, 045108 (2002).

\bibitem{Feldman}
B.E.~Feldman, J.~Martin, and A.~Yacoby, Nat. Phys. {\bf 5}, 889 (2009).

\bibitem{Zhao}
Y.~Zhao, P.~Cadden-Zimansky, Z.~Jiang, and P.~Kim, Phys. Rev. Lett. {\bf 104}, 066801 (2010).

\bibitem{Weitz}
R.T.~Weitz, M.T.~Allen, B.E.~Feldman, J.~Martin, and A.~Yacoby,
Science {\bf 330}, 812 (2010).

\bibitem{Martin}
J.~Martin, B.E.~Feldman, R.T.~Weitz, M.T.~Allen, and A.~Yacoby,
Phys. Rev. Lett. {\bf 105}, 256806 (2010).

\bibitem{Kim}
S.~Kim, K.~Lee, and E.~Tutuc, Phys. Rev. Lett. {\bf 107}, 016803 (2011).

\bibitem{Freitag}
F.~Freitag, J.~Trbovic, M.~Weiss, and C.~Schonenberger,
Phys. Rev. Lett. {\bf108}, 076602 (2012).

\bibitem{Velasco}
J.~Velasco Jr., L.~Jing, W.~Bao, Y.~Lee, P.~Kratz, V.~Aji, M.~Bockrath, C.~N.~Lau, C.~Varma,
R.~Stillwell, D.~Smirnov, F.~Zhang, J.~Jung, and A.~H.~MacDonald,
Nature Nanotechnology {\bf7}, 156 (2012).

\bibitem{Elferen}
H.~J.~van Elferen, A.~Veligura, E.~V.~Kurganova, U.~Zeitler, J.~C.~Maan, N.~Tombros, I.~J.~Vera-Marun,
and B.~J.~van Wees, 
Phys. Rev. B {\bf 85}, 115408 (2012).

\bibitem{Barlas}
Y.~Barlas, R.~C\^ot\'e, K.~Nomura, and A.H.~MacDonald, 
Phys. Rev. Lett. {\bf 101}, 097601 (2008).

\bibitem{Abergel}
D.~S.~L.~Abergel and T.~Chakraborty, 
Phys. Rev. Lett. {\bf102}, 056807 (2009).

\bibitem{Shizuya}
K.~Shizuya, 
Phys. Rev. B {\bf 79}, 165402 (2009).

\bibitem{Nakamura}
M.~Nakamura, E.V.~Castro, and B.~Dora, 
Phys. Rev. Lett. {\bf 103}, 266804 (2009).

\bibitem{Nandkishore1}
R.~Nandkishore and L.~Levitov, 
arXiv:0907.5395v1; 
Phys. Scr. T {\bf 146}, 014011 (2012).

\bibitem{GGM1}
E.V.~Gorbar, V.P.~Gusynin, and V.A.~Miransky, 
JETP Lett. {\bf 91}, 314 (2010).

\bibitem{GGM2}
E.V.~Gorbar, V.P.~Gusynin, and V.A.~Miransky, 
Phys. Rev. B {\bf 81}, 155451 (2010).

\bibitem{Nandkishore2}
R.~Nandkishore and L.~Levitov, 
arXiv:1002.1966v1. 

\bibitem{Toke}
C.~T\H{o}ke and V.~I.~Fal'ko, 
Phys. Rev. B {\bf83}, 115455 (2011).

\bibitem{Kharitonov1}
M.~Kharitonov, 
arXiv:1105.5386v1. 

\bibitem{GGJM}
E.V.~Gorbar, V.P.~Gusynin, Junji Jia, and V.A.~Miransky, 
Phys. Rev. B {\bf 84}, 235449 (2011).

\bibitem{Jiang2007}
Z.~Jiang, Y.~Zhang, H.L.~St\"ormer, and P.~Kim,
Phys. Rev. Lett. {\bf 99}, 106802 (2007).

\bibitem{footnote}
A possibility of a strong mixing of the Landau levels in bilayer graphene was pointed 
out in Ref.~\onlinecite{Abergel}, where the diagonalization of a many-body Hamiltonian 
with the Coulomb interaction was performed by using the noninteracting many body 
basis in the Hilbert space constructed from the single-particle states.

\bibitem{QHF}
K.~Nomura and A.~H.~MacDonald, Phys. Rev.
Lett. {\bf 96}, 256602 (2006); K.~Yang, S.~Das Sarma, and A.~H.~MacDonald,
Phys. Rev. B {\bf 74}, 075423 (2006);
M.~O.~Goerbig, R.~Moessner, and B.~Dou\c{c}ot, Phys. Rev. B {\bf 74}, 161407(R) (2006);
J.~Alicea and M.~P.~A.~Fisher, Phys. Rev. B {\bf 74}, 075422 (2006);
L.~Sheng, D.~N.~Sheng, F.~D.~M.~Haldane, and L.~Balents, Phys. Rev. Lett. {\bf 99}, 196802 (2007).

\bibitem{MC}
V.~P.~Gusynin, V.~A.~Miransky, S.~G.~Sharapov, and I.~A.~Shovkovy,
Phys. Rev. B {\bf 74}, 195429 (2006);
I.~F.~Herbut, Phys. Rev. Lett. {\bf 97}, 146401 (2006);
Phys. Rev. B {\bf 75}, 165411 (2007);
J.-N.~Fuchs and P.~Lederer, Phys. Rev. Lett. {\bf 98}, 016803 (2007);
M.~Ezawa, J. Phys. Soc. Jpn. {\bf 76} (2007) 094701.

\bibitem{Yang}
K.~Yang, 
Solid State Communications {\bf 143}, 27 (2007).

\bibitem{GGMS08}
E.~V.~Gorbar, V.P.~Gusynin, V.~A.~Miransky, and I.~A.~Shovkovy,
Phys. Rev. B {\bf 78}, 085437 (2008).

\bibitem{Goerbig}
M.O.~Goerbig, 
Rev. Mod. Phys. {\bf 83}, 1193 (2011).

\bibitem{BYM}
Y.~Barlas, K.~Yang, and A.H.~MacDonald, 
Nanotechnology {\bf 23}, 052001 (2012).

\bibitem{Semenoff}
G.~W.~Semenoff and F.~Zhou,
J. High Energy Phys. {\bf 1107}, 037 (2011).

\bibitem{Kharitonov2}
M.~Kharitonov,
Phys. Rev. B {\bf 85}, 155439 (2012).

\bibitem{GGMS11.05}
E.V.~Gorbar, V.P.~Gusynin, V.A.~Miransky, and I.A.~Shovkovy,
 Phys. Scr. T {\bf 146}, 014018 (2012).

\bibitem{Potentials}
Let us emphasize that the Hartree contribution in the gap equation (\ref{SD-equation})
is given in terms of the bare interlayer potential, while the Fock contribution is expressed in terms
of the full potential.

\bibitem{footnote2}
Strictly speaking, the terms proportional to the charge density $\mbox{tr}[G(0;0)]$
should be corrected by adding the contributions from all charges in the system, including the ``background''
ones. Because of the overall neutrality of a bilayer device, therefore, such terms must vanish even away from
the neutrality point (for more detail, see a discussion of this issue in Ref.~\onlinecite{GGJM}).

\bibitem{Hald} F.~D.~M.~Haldane, 
Phys. Rev. Lett. {\bf 61}, 2015 (1988).

\bibitem{footnote3} There are ${8\choose 4} =70$ different solutions for the 
$\nu = 0$ QH state in this model.\cite{GGJM} However, depending on the values of 
external magnetic and electric fields, only the SP solution or the LP one is
the ground state in the present approximation.

\bibitem{collective} As shown in Refs.~\onlinecite{Shizuya,Roldan,Losovik}, 
the Landau level mixing can also significantly change the properties of collective excitations in graphene.

\bibitem{Roldan} 
R.~Rold\'an, J.-N.~Fuchs, and M.~O.~Goerbig, 
Phys. Rev. B {\bf 80}, 085408 (2009);
Phys. Rev. B {\bf 82}, 205418 (2010); 
R.~Rold\'an, M.~O.~Goerbig, and  J.-N.~Fuchs, 
Semicond. Sci. Technol. {\bf 25}, 034005 (2010).

\bibitem{Losovik} 
Yu.~E.~Lozovik and A.~A.~Sokolik, 
arXiv:1111.1176. 

\bibitem{Mayorov}
A.~S.~Mayorov, D.~C.~Elias, M.~Mucha-Kruczynski, R.~V.~Gorbachev, T.~Tudorovskiy,
A.~Zhukov, S.~V.~Morozov, M.~I.~Katsnelson, V.~I.~Fal'ko, A.~K.~Geim, and K.~S.~Novoselov,
Science {\bf 133}, 860 (2011).

\bibitem{Cserti} G.~D\'{a}vid, P.~Rakyta, L.~Oroszl\'{a}ny,
and J.~Cserti, Phys. Rev. B {\bf 85}, 041402(R) (2012).

\bibitem{Cote} 
R.~C\^ot\'e, J.~Lambert, Y.~Barlas, and A.~H.~MacDonald,
Phys. Rev. B {\bf 82}, 035445 (2010);
R.~C\^ot\'e, J.~P.~Fouquet, and W.~Luo,
Phys. Rev. B {\bf 84}, 235301 (2011).

\end{thebibliography}
\end{document}